\documentclass[aip, jcp, longbibliography, twocolumn, floatfix, superscriptaddress, reprint, 10pt]{revtex4-1}

\usepackage[english]{babel}
\usepackage[utf8]{inputenc}
\usepackage[paperwidth=210mm,paperheight=297mm,centering,hmargin=2cm,vmargin=2.5cm]{geometry}
\usepackage[utf8]{inputenc}
\usepackage{verbatim}
\usepackage{comment}
\usepackage{bm}
\usepackage{amsmath}
\usepackage{amsfonts}
\usepackage{natbib}
\usepackage{pdfpages}
\usepackage{graphics}
\usepackage{float}
\usepackage{hyperref}
\usepackage[capitalize]{cleveref}
\usepackage{multirow}
\usepackage{booktabs}
\usepackage{makecell}
\usepackage{outlines}
\usepackage{lipsum}
\usepackage{siunitx}
\usepackage[version=4]{mhchem}
\usepackage{pdfpages}
\usepackage{pgffor}
\usepackage{xcolor}
\usepackage{ulem}
\ifdefined\DIRACREP
\else
\newcommand{\DIRACREP}{}

\usepackage{physics}
\usepackage{xparse}
\ifdefined\COSMOMATHS
\else
\newcommand{\COSMOMATHS}{}

\newcommand{\mbf}[1]{\ensuremath{\mathbf{#1}}}

\newcommand{\D}[1]{\operatorname{d}{\!#1}\,}

\fi %

\NewDocumentCommand{\rep}{s d<| d|>}{%
\IfBooleanTF{#1}{
   \IfValueTF{#2}{
       \IfValueTF{#3}{\braket{#2}{#3}}{\bra{#2}}
       }{
       \IfValueTF{#3}{\ket{#3}}{}
       }
   }{
   \IfValueTF{#2}{
       \IfValueTF{#3}{\braket*{#2}{#3}}{\bra*{#2}}
       }{
       \IfValueTF{#3}{\ket*{#3}}{}
       }
   }
}

\NewDocumentCommand{\rbra}{sm}{\IfBooleanTF{#1}{\rep*<#2|}{\rep<#2|}}
\NewDocumentCommand{\rket}{sm}{\IfBooleanTF{#1}{\rep*|#2>}{\rep|#2>}}
\NewDocumentCommand{\rbraket}{smom}{
    \IfBooleanTF{#1}{
        \IfNoValueTF{#3}{\rep*<#2||#4>}{\rep*<#2|#3\rep*|#4>}
    }{
        \IfNoValueTF{#3}{\rep<#2||#4>}{\rep<#2|#3\rep|#4>}
    }
}

\NewDocumentCommand{\field}{o m e{_} e{^} o e{_} e{^}}{
\IfValueTF{#5}{\overline{
  #2\IfValueT{#3}{_#3}\IfValueT{#4}{^{\otimes #4}} %
  \otimes
  #5\IfValueT{#6}{_#6}\IfValueT{#7}{^{\otimes #7}} %
  \IfValueT{#1}{;#1}
}}{
  \IfValueTF{#4}{\overline{
     #2\IfValueT{#3}{_#3}\IfValueT{#4}{^{\otimes #4}}
     \IfValueT{#1}{;#1}
  }}
  {#2\IfValueT{#3}{_#3}}
}
}

\NewDocumentCommand{\frho}{o e{_} e{^}}{
\field[#1]{\rho}_{#2}^{#3}
}

\newcommand{\br}{\mbf{r}}
\newcommand{\bx}{\mbf{x}}
\newcommand{\bxhat}{\hat{\mbf{x}}}

\newcommand{\e}{a}  %

\NewDocumentCommand{\ex}{e_}{
\IfValueTF{#1}{\e_{#1}\bx_{#1}}{\e\bx}
}  %

\NewDocumentCommand{\lm}{e_}{
\IfValueTF{#1}{l_{#1}m_{#1}}{lm}
}
\NewDocumentCommand{\nlm}{e_}{
\IfValueTF{#1}{n_{#1}\lm_{#1}}{n\lm}
}
\NewDocumentCommand{\enlm}{e_}{
\IfValueTF{#1}{\e_{#1}\nlm_{#1}}{\e\nlm}
}
\NewDocumentCommand{\en}{e_}{
\IfValueTF{#1}{\e_{#1}n_{#1}}{\e n}
}
\NewDocumentCommand{\nlk}{e_}{
\IfValueTF{#1}{n_{#1}l_{#1}k_{#1}}{nlk}
}
\NewDocumentCommand{\enlk}{e_}{
\IfValueTF{#1}{\e_{#1}\nlk_{#1}}{\e\nlk}
}
\NewDocumentCommand{\enl}{e_}{
\IfValueTF{#1}{\en_{#1}l_#1}{\en l}
}

\NewDocumentCommand{\nnl}{s}{
\IfBooleanTF{#1}{n_1 n_2 l}{n_1; n_2; l}
}
\NewDocumentCommand{\ennl}{s}{
\IfBooleanTF{#1}{\en_1 \en_2 l}{\en_1; \en_2; l}
}

\NewDocumentCommand{\gslm}{s}{
\IfBooleanTF{#1}{\sigma\lambda\mu}{\sigma;\lambda\mu}
}

\fi %

\ifdefined\COSMOMODELS
\else
\newcommand{\COSMOMODELS}{}
\usepackage{bm,upgreek}

\newcommand{\feat}{\upxi}
\newcommand{\bfeat}[0]{\ensuremath{\bm{\upxi}}}

\fi %

\newcommand{\heapot}{HEA25-4-NN}
\newcommand{\heatf}{HEA$_\text{all}$}

\newcommand{\mc}[1]{{\color{blue}{#1}}}

\newcommand{\SM}{Supplemental Material\cite{smat}}
\newcommand{\Ualchemical}{\mathbf{u}_\text{alch}}
\newcommand{\Nalchemical}{n_\text{alch}}

\newcommand{\rev}[1]{{ #1}}

\makeatletter
\AtBeginDocument{\let\LS@rot\@undefined}
\makeatother

\begin{document}

\setcitestyle{super}

\title{Modeling high-entropy transition-metal alloys with alchemical compression}
\author{Nataliya Lopanitsyna}
\author{Guillaume Fraux}
\affiliation{Laboratory of Computational Science and Modeling, Institute of Materials, \'Ecole Polytechnique F\'ed\'erale de Lausanne, 1015 Lausanne, Switzerland}
\author{Maximilian A.~Springer}
\author{Sandip De}
\affiliation{BASF SE, Carl-Bosch-Stra{\ss}e 38, 67056 Ludwigshafen, Germany}
\author{Michele Ceriotti}
\affiliation{Laboratory of Computational Science and Modeling, Institute of Materials, \'Ecole Polytechnique F\'ed\'erale de Lausanne, 1015 Lausanne, Switzerland}

\onecolumngrid
\begin{abstract}

Alloys composed of several elements in roughly equimolar composition, often referred to as high-entropy alloys, have long been of interest for their thermodynamics and peculiar mechanical properties, and more recently for their potential application in catalysis.
They are a considerable challenge to traditional atomistic modeling, and also to data-driven potentials that for the most part have memory footprint, computational effort and data requirements which scale poorly with the number of elements included.
We apply a recently proposed scheme to compress chemical information in a lower-dimensional space, which reduces dramatically the cost of the model with negligible loss of  accuracy, to build a potential that can describe 25 \emph{d}-block transition metals. The model shows semi-quantitative accuracy for prototypical alloys, and is remarkably stable when extrapolating to structures outside its training set. 
We use this framework to study element segregation in a computational experiment that simulates an equimolar alloy of all 25 elements, mimicking the seminal experiments by Cantor et al., and use our observations on the short-range order relations between the elements to define a data-driven set of Hume-Rothery rules that can serve as guidance for alloy design. We conclude with a study of three prototypical alloys, \ce{CoCrFeMnNi}, \ce{CoCrFeMoNi} and \ce{IrPdPtRhRu}, determining their stability and the short-range order behavior of their constituents. 

\end{abstract}
\twocolumngrid

\maketitle

\section{Introduction}

Almost 20 years have passed since independent work from the groups of Yeh\cite{yeh+04aem} and Cantor\cite{cant+04msea} showed that mixing up to 20 metallic elements in roughly equal parts leads to a smaller-than-expected number of distinct phases, with some corresponding to disordered solid solutions of 4-6 elements. 
These so-called high-entropy alloys (HEAs) have since become the subject of intense study.\cite{cant21pms} On a fundamental level, the observation of the existence of an extended single-phase stability region for alloys with multiple principal components was surprising, and from a technological standpoint it opened up the possibility of designing new materials that defy the limitations of conventional metallurgy and alloy engineering.\cite{geor+20am,li+21pms} %

Besides their metallurgical and mechanical applications, HEAs have been found to be promising catalysts\cite{Sun2021,Wang2021}, especially in electrocatalysis\cite{Huo2021,Zhang2021,Huo2022}. They can efficiently reduce overpotentials and boost activities for, e.g., water splitting\cite{Zhang2018,Bondesgaard2019,Glasscott2019,Jin2019,Lacey2019,Liu2019,Qiu2019,Qiu2019a,Gao2020,Huang2020,Wu2020}, the oxygen reduction reaction \cite{Chen2015,Lffler2018,Lacey2019,Qiu2019a,Li2020}, or the methanol oxidation reaction\cite{Barranco2008,Tsai2009,Wang2014,Chen2015,Yusenko2017} while exhibiting very good stability under reaction conditions. 
These unusual properties are linked to their multi-elemental character, which gives rise to four core effects\cite{Yeh2013,Pickering2016}: the entropy,  'sluggish diffusion', lattice distortion and 'cocktail effect'. 
While the former two enhance the stability, the latter two can explain the high activity in catalysis. First, lattice distortions occur due to atoms being surrounded by atoms of many different atomic radii leading to stress and strain. This alters the electronic structure of the alloy. 
For example, the water splitting activity of a family of AlNiCoIrX (X~=~Mo, Cr, Cu, Nb, V) is superior to \ce{IrO2} because the lattice distortion leads to shorter Ir-O bonds\cite{Jin2019}. Second, the 'cocktail effect' describes unexpected, synergistic effects of the chosen composition. For instance, the non-noble metal HEA CoCrFeMoNi shows activity for the oxygen reduction reaction similar to that of Pt.

From the computational perspective, modeling HEAs poses a number of distinct challenges.  The presence of multiple components requires relatively large simulation cells to unveil microstructures or order-disorder behaviour, while the sluggish diffusion requires long time scales and accelerated sampling techniques to overcome free-energy barriers to atom diffusion. 
Chemical complexity makes empirical forcefields inaccurate, and sampling issues make explicit electronic-structure calculations prohibitively demanding.
As a consequence, the study of HEAs usually relies on on-site cluster expansions\cite{sanc+84pa,nata+21jped}, together with analytical models that allow to capture the qualitative thermodynamic behavior\cite{rao-curt22am}, even though entropic effects beyond configurational ones are known to play an important role.\cite{ma+15am}
More recently, forcefields based on machine learning (ML) have emerged as an alternative approach, allowing to match the accuracy of first-principles calculations, while describing off-lattice distortions and thermal fluctuations\cite{Farkas2018,Byggmstar2021,rose+21npjcm,zhou+22prb}.
However, the majority of ML frameworks for materials modeling exhibit a poor scaling of memory, computation and data requirements with the number of chemical species, and so simulations this far have been restricted to a specific combination of 4-5 elements.
In this paper we introduce a general-purpose ML model for the study of bulk HEAs, that uses a recently-proposed strategy to reduce the dimensionality of chemical space, allowing us to generate an accurate and transferable ML potential that can describe arbitrary mixtures of 25 transition metals. 
The functional form of the model lends itself to an intuitive interpretation of the relations between different transition metals, and careful validation shows that it is capable of accuracy comparable to that of electronic-structure methods in several reference calculations despite the breadth of chemical space it covers. 
We use this potential to reproduce computationally the seminal Cantor experiments on the decomposition of multi-element mixtures, and find a qualitative behavior in the affinity between different species that is consistent with well-known HEAs, allowing us to introduce a data-driven version of the Hume-Rothery rules to guide alloy design. 
We conclude by studying three alloy compositions - the prototypical Cantor alloy CoCrFeMnNi, its Mn$\rightarrow$Mo counterpart that has enhanced catalytic performance, and PdPtIrRuRh -- another promising composition for catalysis. 
In all cases we observe a tendency to phase-separate at low temperature, and that the short-range order observed in high-temperature conditions is indicative of the thermodynamic drive to de-mix.

\section{Alchemical compression of ML representations}

\begin{figure}[btp]
    \centering
    \includegraphics[width=1.0\columnwidth]{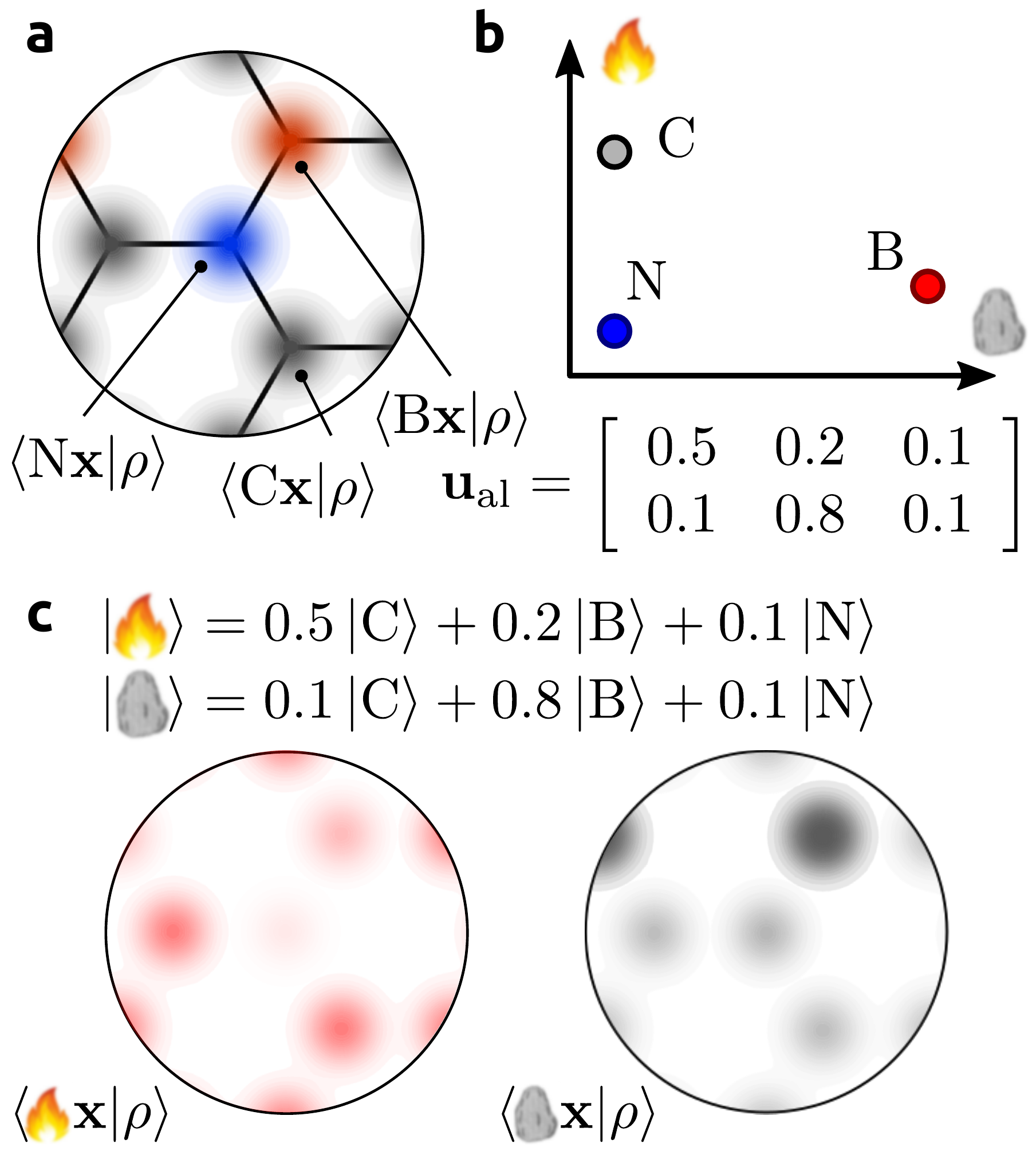}
    \caption{Different interpretations of the alchemical compression scheme. (a) In a conventional density-correlation ML scheme, each type of atoms is associated with a separate density. (b) The entries in the alchemical compression matrix $\Ualchemical$ can be interpreted as describing the ``character'' of each physical element in terms of $\Nalchemical$ pseudoelements - a concept that is not dissimilar from the notion of ``classical elements''.  (c) The structure can be also seen as described in terms of a density of pseudo-elements, for which each site contains a contribuition from each of the compressed channels. 
    }
    \label{fig:alchemy-scheme}
\end{figure}

We follow the approach introduced in Ref.~\citenum{will+18pccp} to reduce the computation, memory and data requirements of a ML model for a chemically-diverse problem.  Here we only give a brief overview, to highlight the key ideas and introduce the notation. The framework relies on the atom-centered density correlation framework\cite{will+19jcp}, which encompasses most of the widespread descriptors for atomic-scale ML, and that is essentially equivalent to the moment tensor potentials\cite{shap16mms} and the atomic cluster expansion\cite{drau19prb}. The reader is invited to read Ref.~\citenum{musi+21cr} (especially Sections 3 and 7.3) for a more pedagogic discussion. The essential ingredient in this framework is the expansion of the neighbor density within an environment $A_i$, that describe the atoms in structure $A$ within a spherical region centered on the $i$-th atom (Fig.~\ref{fig:alchemy-scheme}a, on a basis of radial functions $R_{nl}$ and spherical harmonics $Y_l^m$:
\begin{equation}
\begin{split}
 \rep<anlm||\frho_i>=&
\int \D{\bx} R_{nl}(x) Y_l^m(\bxhat) \rep<a\bx||\frho_i>, \\
\rep<a\bx||\frho_i>\equiv& \sum_{j\in A_i} \delta_{aa_j} g(\bx-\br_{ji}).
\end{split}
\end{equation}
In this expression, $j$ runs over the neighbors of atom $i$, $g(\bx)$ is a Gaussian function (or its Dirac-$\delta$ limit), $\br_{ji}\equiv \br_j - \br_i$ is the interatomic distance vector between the $j$-th and $i$-th atom, $R_{nl}$ enumerates the radial functions and $Y_l^m$ the spherical harmonics. The ket $\rep|\rho_i>$ indicates the $i$-centered neighbor density and the $a$ index identifies the chemical nature of the atoms.

The bra-ket notation serves to emphasize the fact that the discrete coefficients are simply a projection on a basis of the very same quantity as the real-space neighbor density.
We also express the density coefficients with the alternative notation $\rep<an||\frho[\lambda\mu]_i^1>\equiv \rep<an\lambda\mu||\rho_i>$, in which $\rho_i^{\otimes \nu}$ indicates we are describing the $\nu$-neighbors density correlations, and the angular indices $(\lambda,\mu)$ are moved to the ket to highlight that they determine the symmetry of the coefficients with respect to rotations, that is crucial when building equivariant models and when combining density coefficients to evaluate higher-order correlations.

In this work we will use the pair invariants ($\nu=1$, $\lambda=0$, i.e. $\rep<an||\frho_i^1> \equiv \rep<an||\frho[00]_i^1>$), as well as the two-neighbors invariant terms ($\nu=2$, $\lambda=0$). The two-neighbors invariants --- equivalent to SOAP features\cite{bart+13prb} and closely-related to three-body Behler-Parrinello symmetry functions\cite{behl11jcp} --- can be computed as:
\begin{multline}
\rep<a_1 n_1; a_2 n_2; l||\frho_i^2> \propto
\sum_m \rep<a_1 n_1||\frho[lm]_i^1>  \\ \times \rep<a_2 n_2||\frho[lm]_i^1>.\label{eq:soap-ps}
\end{multline}
For readers familiar with the notation used in the SOAP literature\cite{bart+13prb}, the expansion coefficients of the density are often written as $c^a_{nlm}$ and the power spectrum, corresponding to the two-point correlations~\eqref{eq:soap-ps}, as $p_{n_1 n_2 l}^{a_1 a_2}$.
In both forms, it is clear that the number of components one has to consider grows quadratically with the number of species, each element being considered independently in the neighbor density. The generalization to higher-$\nu$ correlations leads to an even steeper increase, but for the dataset we consider here, the computational cost is prohibitive even for two-neighbors correlations.

The key insight in Ref.~\citenum{will+18pccp} is that it is unnecessary --- and possibly detrimental --- to consider elements as independent. 
Similarities in the behavior of elements have inspired the construction in the periodic table,\cite{scha19jpcl} and are routinely used to inform materials design and optimization.
Instead, elements should be mapped to a continuous $\Nalchemical$-dimensional space, where each chemical species is mapped to $\Nalchemical$ \emph{pseudo-species} with a set of coupling coefficients $\Ualchemical$.
Then, the density coefficients can be contracted as
\NewDocumentCommand{\frhot}{o e{_} e{^}}{
\field[#1]{\tilde{\rho}}_{#2}^{#3}
}
\begin{equation}
\rep<bn||\frhot[\lambda\mu]_i^1> \equiv \sum_a u_{ba} \rep<an||\frho[\lambda\mu]_i^1>,\label{eq:alch-cmprss}
\end{equation}
where we use $\tilde{\rho}$ to indicate the alchemically-compressed neighbor density (Fig.~\ref{fig:alchemy-scheme}). 
We note that similar ideas were applied -- without optimizing the contraction coefficients -- in the context of atom-centered symmetry functions\cite{artr+17prb,gast+18jcp}, and that a systematic, rather than data-driven, compression has also been recently applied to a 8-element alloy system in the context of atomic cluster expansion potentials\cite{darb+22npjcm}.
Moreover, there is a large design space of variations on a theme: separate coupling coefficients could be used depending on angular ($\lambda$) and/or radial ($n$) channel, and it would be possible to jointly contract over chemical and radial components -- which was shown to be effective in reducing the number of features with minimal information loss\cite{gosc+21jcp}.
Here we do not explore this design space, because, as we shall see, the pure alchemical contraction appears to be both effective and easy to interpret.
Using these compressed density coefficients~\eqref{eq:alch-cmprss} one can evaluate correlation functions with a cost that still scales exponentially with $\nu$, but with a more benign base, or perform further iterative contraction steps as in Ref.~\citenum{niga+20jcp}. %

To conclude this overview, we note that the alchemical coefficients $\Ualchemical$ enter the expression for the $\nu=2$ features in a quadratic fashion, and so they cannot be directly determined using linear algebra, even if one uses a linear model based on the contracted features.
In Ref.~\citenum{will+18pccp} this issue was tackled with an iterative strategy, alternating a solution of the linear problem with fixed  $\Ualchemical$ and a gradient descent on the coupling coefficients. In the present work, instead, we implement the model using the PyTorch framework\cite{pasz+19nips}, allowing us to use automatic differentiation and gradient descent to optimize simultaneously $\Ualchemical$ and the model weights.

\section{Computational details}

We provide a concise summary of the details of the calculations we perform in this work, \rev{covering the reference electronic-structure calculations, the construction of the training set, the architecture of the ML model, as well as the details of the sampling protocol that we use for simulations in Sections~\ref{sec:cantor} and~\ref{sec:app}.} In the \SM{} we provide representative examples of the typical simulation setup, and additional convergence tests. 

\subsection{Electronic-structure details}
All the reference energies and forces are computed using density-functional theory (DFT), as implemented in the VASP code \cite{VASP}, with the PBESol exchange-correlation functional  \cite{csonka2009assessing}. The core electrons are treated implicitly using projector augmented wave (PAW) pseudopotentials \cite{kresse1999ultrasoft}.
We choose conservative values for the convergence parameters of the electronic structure calculation (see the \SM{} for details): the wave function is expanded in plane waves with a cutoff energy of 550 eV, and the Brillouin zone sampling uses a $\Gamma$ centered Monkhorst-Pack scheme \cite{monk-pack76prb} with an interval between k-points along reciprocal lattice vector 0.04 $\pi$ \AA$^{-1}$.
Even though transition metals often exhibit magnetism, either in the pure phases or in alloys, we perform all our calculations without spin polarization. Even disregarding the fact that ML models that can deal with magnetism are still at a very early stage\cite{novi+22npjcm}, one should consider that we aim to cover a broad chemical range, that includes materials which require different types of approaches to describe accurately their magnetic behavior - band magnetism within the local spin density approximation,\cite{wang+82prb} non-colinear magnetism,\cite{kubl+88jpf} Hubbard-U calculations\cite{kuli+06prl}, etc.
This makes non-polarized calculations a reasonable approximation within the scope of the present work (see also the \SM{}), \rev{even though this limits the accuracy of our reference and our model for magnetic systems - which for example would not be able to predict the stabilization of \emph{bcc} iron over the close-packed polymorphs.}

\subsection{Training set construction}
We generated an original dataset including 25 \emph{d}-block elements, i.e. all transition metals excluding those that are not listed in Ref.~\citenum{li2018combinatorial} as relevant for HEAs (Tc, Cd, Re, Os, Hg).
We generate a total of 25 thousand structures, following a protocol that ensures quasi-random sampling of this high dimensional phase space. We created four subsets of structures based on \emph{bcc} and \emph{fcc} lattices containing 36 or 48 atoms, respectively. All lattice parameters are defined by the average atomic volume of the elements in a structure and scaled up or down by up to 10\% at random to simulate compression and expansion.
The structures in the first three classes include from 3 to 8 randomly selected elements, and in the fourth -- from 3 to 25. In the first class, we included only perfect crystal structures, with random compositions. For the three remaining classes, we shuffled atomic positions around their ideal lattice sites (using a Gaussian distribution of atomic displacement with a standard deviation of 0.2 $\AA$ in the second and fourth classes, and 0.5 $\AA$ in the third), to incorporate the information about interactions in crystals at finite temperatures.

For every class of structures, we generated 100'000 random configurations and selected around 7'000 of the most diverse from every subset using Farthest Point Sampling (FPS)\cite{imba+18jcp} in radial spectrum feature space. 

\subsection{Machine-learning model}

We build ML models based on density-correlation representations, combining an atomic-energy baseline, ridge regression based on pair and 3-body correlation features, and a multi-layer perceptron\cite{hayk94book} based on the 3-body features. Here we discuss briefly the functional form of the different term, and outline the training strategy we followed.
The atomic-energy baseline is simply a linear model that depends exclusively on the nature of the atom at the centre of each environment, $a_i$
\begin{equation}
V^{(\text{aeb})}(A_i) = w^{(\text{aeb})}_{a_i}.
\end{equation}
Even though we train on atomization energies (and so the large dependency of the atomic energies on the details of the pseudopotentials is not an issue) we still find that $V^{(\text{aeb})}$ captures a large fraction of the target variance, and facilitates learning.
The second term we consider is a set of pair energies. We use a Gaussian width of 0.25\AA{}, a cutoff of 6\AA{} and radial scaling following Ref.~\citenum{will+18pccp}; \rev{we expand the density in spherical harmonics and in 12 radial function, enumerated by the $n$ index, and obtained by orthogonalizing Gaussian-type orbitals that cover the range of distances up to the cutoff radius (see e.g. Ref.~\citenum{musi+21jcp} for a precise definition). }
We use different weights depending on the nature of the two atoms, so that in practice the contribution to the potential reads
\begin{equation}
V^{(\text{2B})}(A_i) = \sum_{an} w^{(\text{2B})}_{a_i a n} \rep<a n||\frho_i^1>.
\end{equation}
The third term involves 3-body correlations (SOAP features), computed on top of alchemically-contracted density coefficients, with a linear model
\begin{equation}
V^{(\text{3B})}(A_i) = \sum_{bnb'n'l} w^{(\text{3B})}_{bnb'n'l} \rep<b nb' n'l||\frhot_i^2>.
\end{equation}
We use the same set of weights irrespective of the atom type, because in a 3-body descriptor the nature of the central atom is encoded in the density associated with the Gaussian at $r=0$, so that the compression of the dependency of potentials on the central atom type is achieved implicitly and with the same contraction coefficients used for the neighbor density.

Finally, we include a non-linear term that takes the compressed power-spectrum as input, and feeds it into a Behler-Parrinello-style\cite{behl-parr07prl} multi-layer perceptron\cite{hayk94book}.
First, a linear filter projects the power-spectrum features into 80 input neurons, $\bfeat^{(0)}$, to which hyperbolic tangent activation functions are applied. A second linear layer combines the outputs of the neurons, feeding them to one hidden layer of the same size. Finally, the outputs are linearly combined to yield the atomic energy
\begin{equation}
\begin{split}
\feat_q^{(0)}(A_i) = & \sum_{bnb'n'l} w^{(NN,0)}_{qbnb'n'l} \rep<b nb' n'l||\frhot_i^2>, \\
V^{(\text{NN})}(A_i) = &F(\bfeat^{(0)}(A_i))
\end{split}
\end{equation}
We use this simple neural network --- built on top of the compressed power-spectrum features --- because we want a simple and well-understood term that can incorporate non-linearity without exploding the design space, and because we want to show that our alchemical compression scheme can be readily applied to several well-established ML schemes.
It is possible (and likely) that alternative frameworks, e.g. increasing further the body order,  may allow for a better-performing model, but as we shall see this approach is sufficient to achieve state-of-the-art accuracy together with a stable and interpretable model.

The parameters of $V^{(\text{3B})}$ and $V^{(\text{NN})}$ implicitly include the alchemical coupling matrix $\Ualchemical$; for this reason, we optimize all models with gradient descent, relying on backpropagation as implemented in PyTorch\cite{pasz+19nips}. A ridge penalty term is included on all weights, to reduce the risk of overfitting. We find that (possibly due to the presence of large linear components that contribute a quadratic term to the $L^2$ loss) a deterministic L-BFGS optimizer\cite{liu-noce89mp} performs much better than stochastic gradient descent.

\subsection{Sampling details}
\label{sub:sampling}
Molecular dynamics (MD) is well-suited to describe structural relaxation of the atomic coordinates. However, long-range diffusion in the solid phase occurs through vacancies, and is too slow to be simulated explicitly by MD.
To overcome this time scale problem, we use a combination of techniques to facilitate thorough sampling of atomic ordering. Our base protocol involves performing molecular-dynamics simulations in the constant-temperature/constant-pressure \emph{NpT} ensemble.\cite{ande80jcp}
We use a conservative time step of 2~fs, an isotropic barostat~\cite{buss+09jcp} with a time constant of 200~fs coupled to an  optimal-sampling colored-noise thermostat\cite{ceri+10jctc}, and an aggressive thermostat for the ions, alternating an optimal-sampling Langevin equation with a stochastic velocity rescaling\cite{buss+07jcp} with a time constant of 10~fs.
We accelerate sampling of the compositional (dis)order by performing Monte Carlo steps in which the nature of two atoms in the system is exchanged, with a Metropolis acceptance criterion\cite{metr+53jcp}.
We perform on average one exchange attempt per MD time step. Both the MD and the MC step conserve the Boltzmann distribution (except for a negligible finite time-step error), and so the combined MD/MC protocol is consistent with canonical sampling.
In order to further accelerate sampling, we also use replica exchange molecular dynamics (REMD)\cite{earl-deem05pccp} -- a technique in which multiple trajectories at different temperatures are performed in parallel. Periodically, structures are exchanged between temperatures, using a Monte Carlo procedure that preserves the Boltzmann distribution for each thermodynamic state.
The fact that each trajectory is brought through cycles of heating and annealing accelerates conformational sampling and reduces the correlation time of observables that are associated with activated events at low temperature.
Unless otherwise specified, we use temperature replicas distributed according to a geometric progression between two extremal values $T_\text{min}$ and $T_\text{max}$.
For all MD/MC simulations we use the i-PI universal force engine\cite{kapi+19cpc}, that includes an implementation of element exchange moves\cite{imba-ceri21prm} and a flexible implementation of replica exchange\cite{petr+15jcc}.

\section{Alchemical learning}

As discussed above, the compression scheme in Eq.~\eqref{eq:alch-cmprss} is just one of the many approaches one could take to reduce the dimensionality of the density expansion coefficients.
One of the appealing features of this specific implementation is that it can be interpreted relatively easily, and that it allows us to extract physical-chemical insights through an introspection of the model parameters and performance.

\begin{figure}[bp]
    \centering
    \includegraphics[width=1.0\columnwidth]{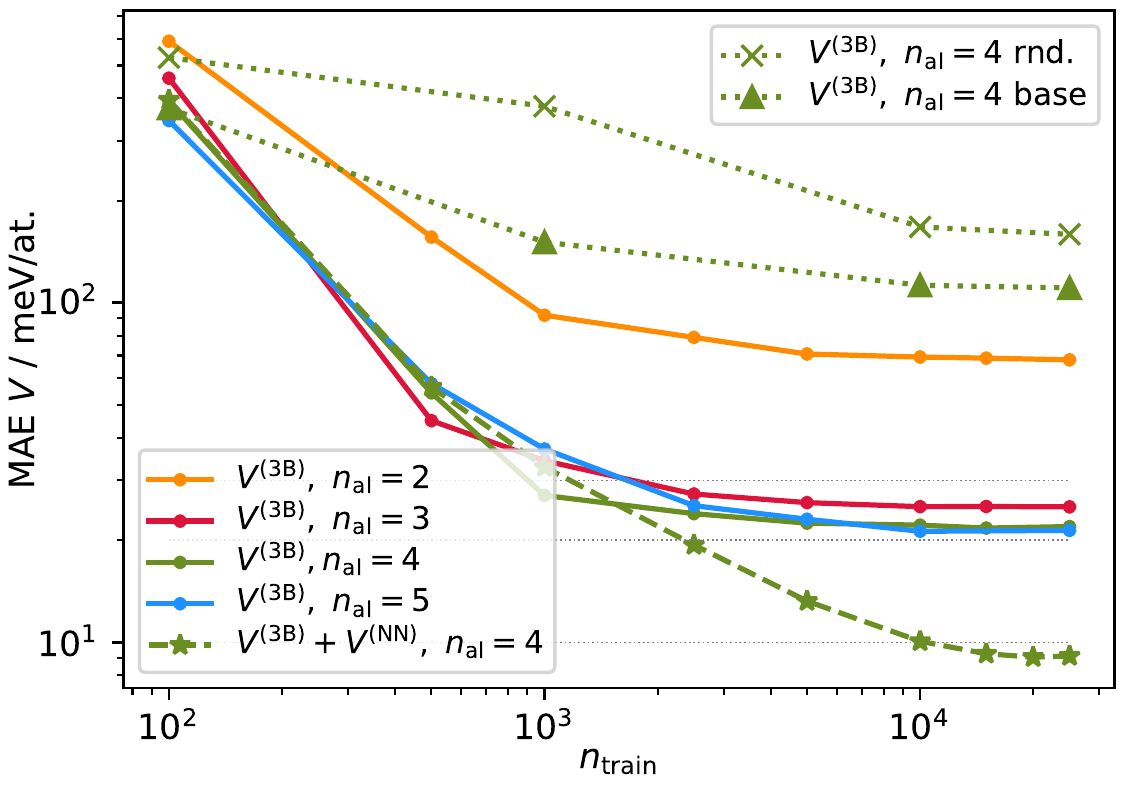}
    \caption{Learning curves for different models. Full lines correspond to models built using only $V^{(\text{aeb})}$ and $V^{(\text{3B})}$, with $\Nalchemical$ pseudo-elements (all optimized iteratively). \rev{The dotted green curves are obtained with a $\Ualchemical$ filled with uniform random numbers (rnd.) and with the weights we use as an initial guess for the optimized models (base), that are built based on physical priors following the scheme discussed in Ref.~\citenum{will+18pccp}}.  The dashed green line corresponds to a model that includes $V^{(\text{aeb})}$ and $V^{(\text{3B})}$, as well as the full set of pair potentials and a non-linear term built on top of the contracted power spectrum features $V^{(\text{NN})}$.
    }
    \label{fig:learning-curves}
\end{figure}

\subsection{Learning curve analysis}

We begin by considering linear models based on contracted power-spectrum features, supplemented by an atomic energy baseline term, $V^{(\text{aeb})}+V^{(\text{3B})}$.
We perform separate training exercises, using only energy as targets, and restricting the alchemical contraction to 2, 3, 4, 5 pseudoelements. \rev{For each model we compute learning curves by converging the loss at a given number of training structures $n_\text{train}$, then increase the train size and continue the optimization restarting from the previous weights. }
Given that the optimization procedure is rather demanding, we do not perform multiple train/test split, but use consistently the same shuffle with up to 25'000 structure used for training and a hold-out set containing 500 configurations used for testing.
Even though the accuracy does depend slightly on the shuffle, and on the initialization of the weights, we find that the qualitative observations we present here are robust. 

Figure~\ref{fig:learning-curves} shows a behavior similar to that observed in Ref.~\citenum{will+18pccp} for an analogous exercise on the elpasolites data set\cite{fabe+16prl}: at the smaller train set sizes a very aggressive compression is effective at obtaining a robust model, but with more training data the learning curves saturate. Increasing the number of pseudo-elements $\Nalchemical$ delays saturation, but the improvement going from $\Nalchemical=3$ to $\Nalchemical=4$ is negligible, and the learning curves for $\Nalchemical=5$ sits almost exactly at the same value. 
This indicates that, from the point of view of 3-body interactions, 3-4 pseudo-elements are sufficient to saturate the descriptive power of a linear model. 
Note that the optimization of $\Ualchemical$ is critical to achieve such efficient compression: a model that uses fixed, random values for the contraction weights, \rev{as well as one that uses a fixed, physically-inspired initialization of $\Ualchemical$,} lead to an order of magnitude increase in the saturation error, even with $\Nalchemical=4$ (Fig.~\ref{fig:learning-curves}).

Given the saturation of $V^{(\text{3B})}$, we proceed to increase the effective body-order of the potential adding a non-linear NN layer on top of the contracted power spectrum, $V^{(\text{NN})}$\rev{, which introduces about $160'000$ additional model parameters, mostly associated with the contraction of the $\rep|\field{\tilde{\rho}}_i^2>$ features to the 80 input features of the NN}. 
Furthermore, we also include a non-compressed two-body potential $V^{(\text{2B})}$, for which we also consider a slightly larger cutoff distance. 
This 2-body term, on its own, does not improve significantly the limiting accuracy of the model (reinforcing the notion that the alchemical contraction is converged) but we include it because it is inexpensive to compute, and has been shown in the past to lead to more stable models, whose performance degrade more gently in the extrapolative regime\cite{deri-csan17prb,deri+21cr}.
Incorporating a non-linear term in the model allows to overcome the saturation of the learning curve (Fig.~\ref{fig:learning-curves}, dashed green line). 
The non-linear $\Nalchemical=4$ model reaches a validation-set mean absolute error (MAE) below 10 meV/atom. We discuss further the accuracy of this model (that we will refer to as the \heapot{}) in Section~\ref{sec:validation}.

\subsection{A 3D periodic table for the transition metals}

The alchemical coupling matrix associates to each of the physical elements a vector of size $\Nalchemical$, that can be regarded as the ``composition'' of that element in terms of a set of pseudo-elements (Fig.~\ref{fig:alchemy-scheme}b). 
Thus, different atomic species can be seen as points in a continuum space, and can be visualized as such to gain insights into the data-driven similarities that arise from the optimization of $\Ualchemical$ to achieve the most accurate regression of the target. 
To make the visualization independent on unitary transformations of the weight matrix, we perform a principal component analysis. 

\begin{figure}[tbhp]
    \centering
    \includegraphics[width=1.0\columnwidth]{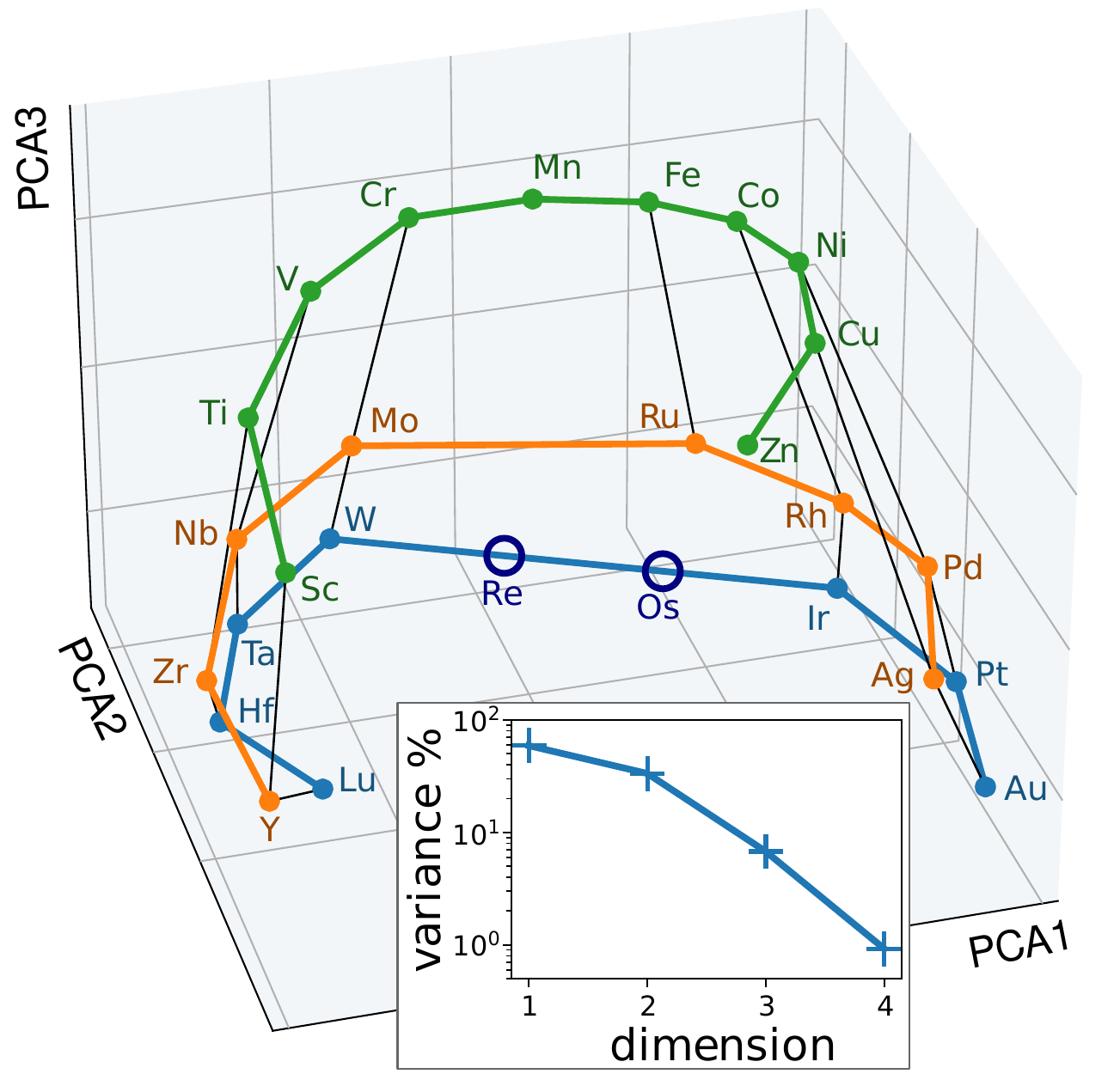}
    \caption{Top-3 principal components of the alchemical coupling matrix $\Ualchemical$ for the \heapot{} model. The periods are highlighted with orange, blue and green lines, and the columns are indicated by black thin lines. 
    Interpolated positions for Re and Os are indicated with empty circles. The inset shows the decay of the explained variance for the four principal components.}
    \label{fig:coupling-weights}
\end{figure}

The eigenvalues of the covariance matrix indicate the magnitude of the various components  (their \emph{explained variance}), and provide another indication of the importance of successive increases in the dimensionality of the alchemical space. 
We observe a quick decrease of the explained variance, with the fourth component typically amounting to less than 2\%{} of the variance (Fig.~\ref{fig:coupling-weights}, inset). 
This confirms that the first three components provide sufficient descriptive power to capture the difference in behavior between transition metals.
We can then look at how the $d$-block elements appear when projected along the top three principal components of $\Ualchemical$ (Fig.~\ref{fig:coupling-weights}). We focus on the weights from the \heapot{} model, but the qualitative features of the alchemical projections are similar also for other models in Fig.~\ref{fig:learning-curves} (see the \SM{}). 
The elements are arranged in a way that is strongly reminiscent of their placement in the $d$ block: the third principal direction corresponds to the period, while the first two dimensions are associated with a semicircular arrangement, with the elements appearing in the same order as the columns in the conventional periodic table. 
Interestingly, this arrangement is reminiscent of that used for the $d$ block in some of the alternative representations of the periodic table, such as the Benfey spiral\cite{seab64book}. It indicates that, from the point of view of the construction of an interatomic potential, zinc is closer to scandium then it is to the atoms in the middle of the transition metals block.

\subsection{Alchemical interpolation}
\label{sec:alch_inter}

The elements we have not considered leave a clear gap in the arrangement of the alchemical coupling weights, and it is interesting to see how accurate a model that places rhenium and osmium between tungsten and iridium fares in predicting their properties without additional fitting. 

We pick 60 structures from the hold-out set, containing distorted configurations with random composition. The MAE for these structures when using the $\Nalchemical=4$ model using only $V^{(\text{aeb})}$ and $V^{(\text{3B})}$ is 13 meV/atom. We then substitute some random atoms with Re and Os, without changing the positions, and re-compute their energies with analogous DFT settings.

We then build a model in which we simply take the parameters optimized for the 25-elements dataset, and complete them by adding atomic-energy baselines for Re and Os (obtained by training on the residual a two-parameter model that depends exclusively on the Re and Os content) and by adding pseudoelement weights that interpolate linearly between W and Ir (see Fig.~\ref{fig:coupling-weights}):
\begin{equation}
u_{b\ce{Re}} = \frac{2}{3} u_{b\ce{W}} + \frac{1}{3} u_{b\ce{Ir}}, \quad 
u_{b\ce{Os}} = \frac{1}{3} u_{b\ce{W}} + \frac{2}{3} u_{b\ce{Ir}}. 
\end{equation}
The powerspectrum model weights are unchanged: we are effectively interpolating in pseudoelement space. 
The the resulting model yields exactly the same predictions for structures that do not contain Os and Re, and has a MAE of only 24 meV/atom for the test structures that include the two species {(see also the \SM{})}. The model is also sufficiently stable to run molecular dynamics simulations for Re and Os containing structures. 

This example underscores the advantages of the interpretable functional form we use to implement alchemical dimensionality reduction. It also opens up the possibility of designing simulation protocols that include smooth ``alchemical transformations'', in a similar spirit as the framework pioneered by von Lilienfeld et al.\cite{sepp+10jcp}, and with some similarities to the virtual crystal approximation that is often used to describe approximately random alloys\cite{bell-vand00prb}. 
For example, one could use thermodynamic integration to compute the change in chemical potential associated with an element substitution by running simulations with a mixed potential, in which the alchemical coupling weights are gradually transformed between the values associated with two elements.

\section{Validation of the potential}
\label{sec:validation}

We now assess the accuracy and stability of the model we use in the rest of this work, which combines a 4-pseudoelement contraction of the powerspectrum with a multi-layer perceptron. 
We aim to provide benchmarks that are easy to reproduce, but that reflect the performance of the model in relevant simulation tasks, and we envisage that any comparative study would include most of these and not only cross-validation statistics. 
To contextualize and provide a reference scale for our results, we report in the \SM{} similar validation results for the general-purpose, universal graph neural network M3GNet\cite{Chen2022}. 
In all cases \heapot{}, which admittedly has a narrower scope of applicability, outperforms M3GNet by a large margin.

\begin{figure}[tbhp]
    \centering
    \includegraphics[width=1.0\columnwidth]{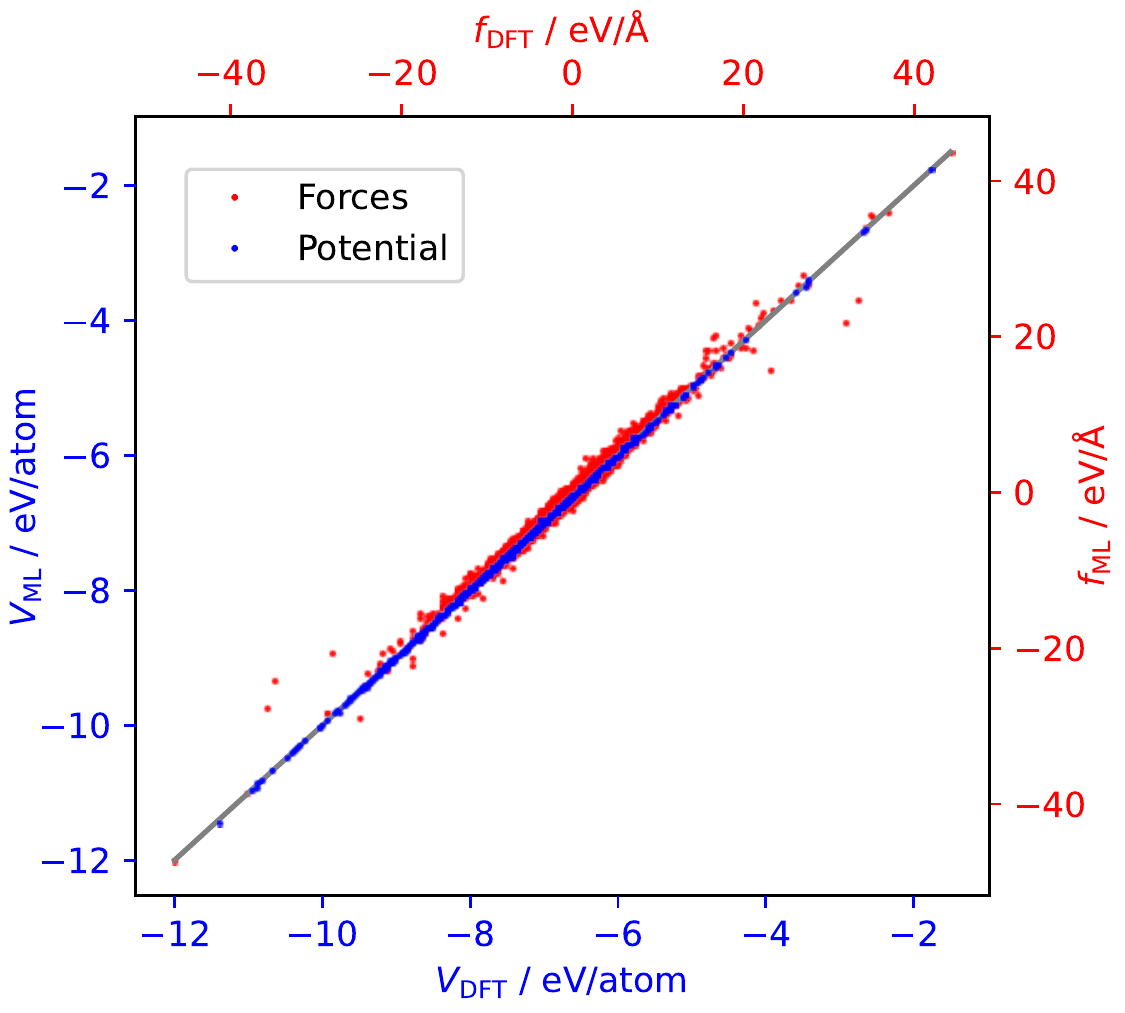}
    \caption{Parity plot between reference energy and forces and the values computed with the \heapot{} model, for a hold-out set of 500 structures, randomly selected from the \rev{overall pool of structures.} Energy error: 10 meV/atom mean absolute errror (MAE), 14 meV/atom root mean square error (RMSE), Force error: 190 meV/\AA{} MAE, 280 meV/\AA{} RMSE. }
    \label{fig:validation-parity}
\end{figure}

\subsection{Hold-out validation of the \heapot{} model}

We train the \heapot{} potential by progressively increasing the train set size, until we run the final optimization on 25'000 structures, including forces for 2'000 of them. We hold out 500 structures and use them for validation. 
The parity plot between targets and predictions demonstrates the accuracy of the model (Fig.~\ref{fig:validation-parity}), which is remarkable given the diversity of the dataset, that contains random combinations of up to 25 elements, and  highly-distorted structures.

\begin{figure}[tbhp]
    \centering
    \includegraphics[width=1.0\columnwidth]{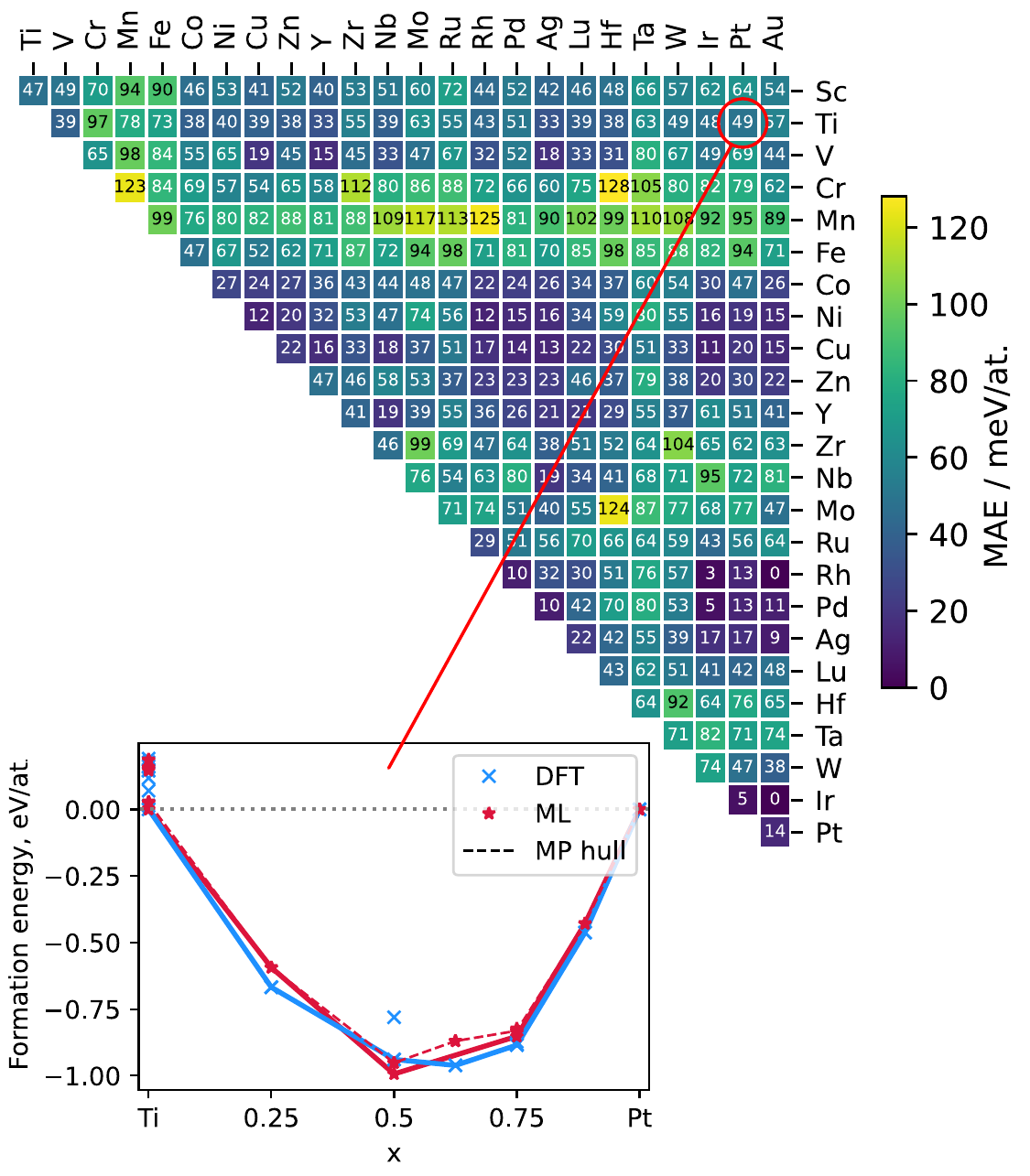}
    \caption{MAE for the formation energy of binary compounds from the Materials Project database.
    The inset shows a representative hull plot for the Ti--Pt system, highlighting the hulls obtained from the single-point DFT calculations and the ML predictions. The dashed line identifies the structures that are stable based on the energies available in the Materials Project database. 
    }
    \label{fig:validation-ch}
\end{figure}

\subsection{Binary convex hulls}
Even though the \heapot{} is clearly geared towards multi-component simulations, it is important that it also provides reasonable results for simpler compositions, as these may appear spontaneously when complex alloys de-mix and form precipitates. 
We collect 1438 binary intermetallic structures out of more than 146k crystal structures from the Materials Project database\cite{jain+13aplm}, and re-compute their energies with single-point calculations using our DFT setup, as well as with the \heapot{} model. 
We discard 23 structures for which our DFT calculations did not converge and 10 that correspond to configurations that are too dissimilar from the bulk structures we consider here (see \SM{}). 
For the remaining structures, the MAE error for the cohesive energy is 62 meV/at. and for the formation energies is 63 meV/at, which is higher than the cross-validation error, but still remarkably accurate for extrapolative predictions. 
It is worth noting that the MAE discrepancy between our DFT calculations and those saved in the MP records is 65 meV/at.; \rev{this is due to the significant difference in the details of the electronic structure calculations, e.g. the use of Hubbard U corrections for some structures in the MP protocol, and neglect of spin polarization in ours.} This observation underscores that the details of the electronic structure calculations can have an impact comparable to the accuracy of our ML model. 
We then use this data to compute binary convex-hull diagrams for all element pairs. In Fig.~\ref{fig:validation-ch} we show a representative example for the Ti--Pt system. The overall shape of the hull is usually well-reproduced, but often \heapot{} predicts different stable polymorphs than DFT, and/or mis-predicts the stability of certain compositions (as it is the case for \ce{TiPt2} in the figure). However, these qualitative errors are usually associated with situations in which a small energy shift can bring a composition above the hull boundary, and even in a fully ab initio study it would not be possible to determine conclusively its thermodynamic stability. The full list of hulls is included in the \SM{}.
Fig.~\ref{fig:validation-ch} also shows an overview of the accuracy of the prediction of formation energies for all phases (stable and unstable) as a function of composition. Errors are not uniform: some elements such as Mn, that have the tendency of forming complex crystal structures, yield larger errors, while others such as Cu or Ni usually yield errors comparable to the validation set. 
It would be trivial to improve the accuracy of the model for binary structures and pure element polymorphs by including this small number of additional structures in the training set. We chose not to do that to avoid introducing biases in the accuracy depending on the different abundance of structures in the MP database.
In the future, we plan to extend systematically our training set to incorporate disordered and liquid structures.

\begin{figure}[tbhp]
    \centering
    \includegraphics[width=1.0\columnwidth]{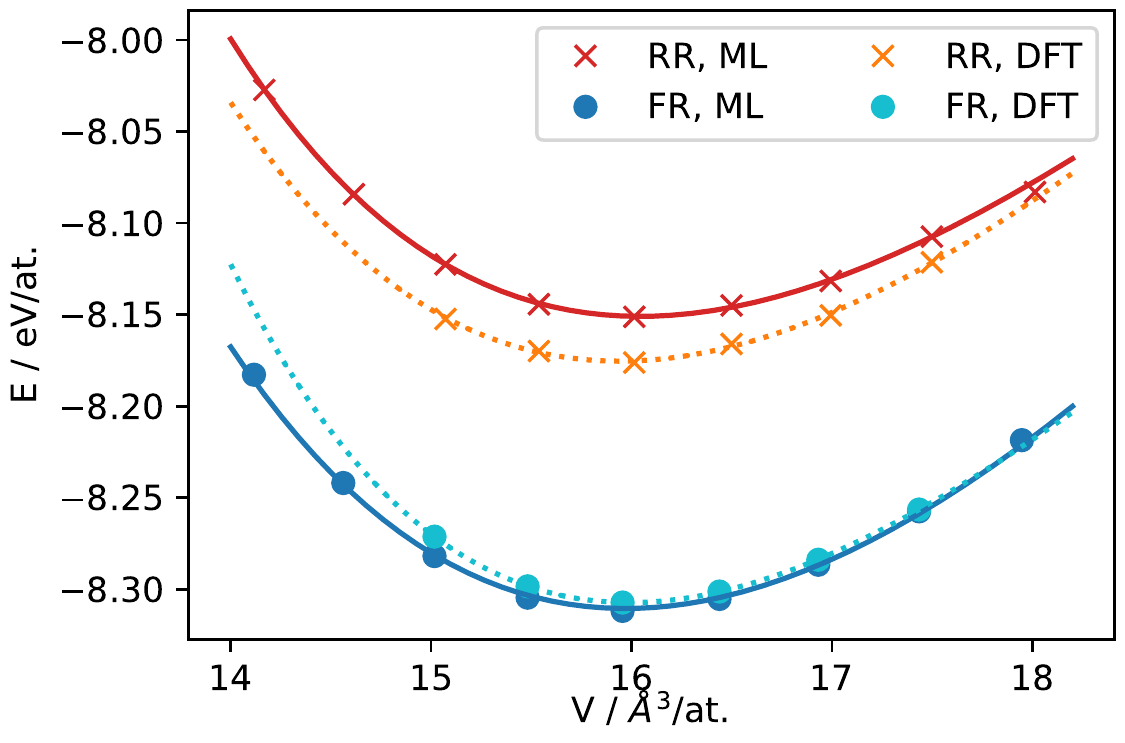}
    
    \renewcommand{\arraystretch}{1.1}
    \begin{tabular}{lcccc}    
    \hline\hline
                        & RR, ML & RR, DFT & FR, ML & FR, DFT \\
    \hline
    $E_0$ (eV/at.)      & -8.151 & -8.175 & -8.310 &  -8.307 \\
    $V_0$ (\AA$^3$/at.) & 16.04  &  15.94 & 15.97  & 16.00    \\ 
    $B_0$ (GPa)         & 132    & 141    & 146    & 162      \\ 
    $B'_0$\ \ \ \ \ \ \ & 6.7    & 6.1    & 4.8    & 8.0      \\ 
    \hline\hline
    \end{tabular}
    \caption{Equation of state for the random relaxed (RR) and fully relaxed (FR) structures (see text for the full definition), computed with the \heapot{} potential and with the reference DFT. Birch-Murnaghan parameters for cohesive energy ($E_0$), equilibrium volume ($V_0$), bulk modulus ($B_0$), bulk modulus derivative ($B'_0$) are given in the table.}
    \label{fig:validation-eos}
\end{figure}

\subsection{Energy and equation of state}

We prepare a $5\times 5 \times 5$ $fcc$ supercell, containing 5 atoms of each of the 25 elements, arranged randomly on the lattice. 
We relax the geometry of the structure, and the volume of the supercell, using the \heapot{} potential. We refer to this structure as the random relaxed (RR) structure.  
Starting from the same configuration, we also perform a slow annealing trajectory, combining molecular dynamics and atom exchange moves, to obtain a structure in which the arrangement of elements is not random, but more energetically favorable. We refer to this structure as the fully-relaxed (FR) structure.
In both cases, the atoms relax away from $fcc$ lattice positions, and the resulting structure within the supercell is rather disordered.
We then introduce an isotropic compression or expansion of the two structures, relaxing the coordinates of the atoms within the cell, and fit a Birch-Murnaghan equation of state to the resulting energy-volume curves. 
We repeat the fixed-cell relaxation with the reference DFT, and compare the resulting equations of state (Fig.~\ref{fig:validation-eos}). The error on the cohesive energy $E_0$ is comparable to the test error (24meV for $E_0^{(\text{RR})}$, 3meV for $E_0^{(\text{FR})}$), and much smaller than the energy gain associated with the annealing of the lattice occupations (\rev{$E_0^{(\text{RR})}-E_0^{(\text{FR})}$} is about 150 meV/atom), indicating that \heapot{} is reliable for assessing the energetics of ordering in a random alloy. 
The equilibrium volume and bulk modulus for the two structures are also in good agreement, with errors below 1\%{} and 10 \%{}, respectively -- comparable with the typical discrepancy between different DFT approximations or between DFT and experiments. 

\begin{figure}[tbhp]
    \centering
    \includegraphics[width=1.0\columnwidth]{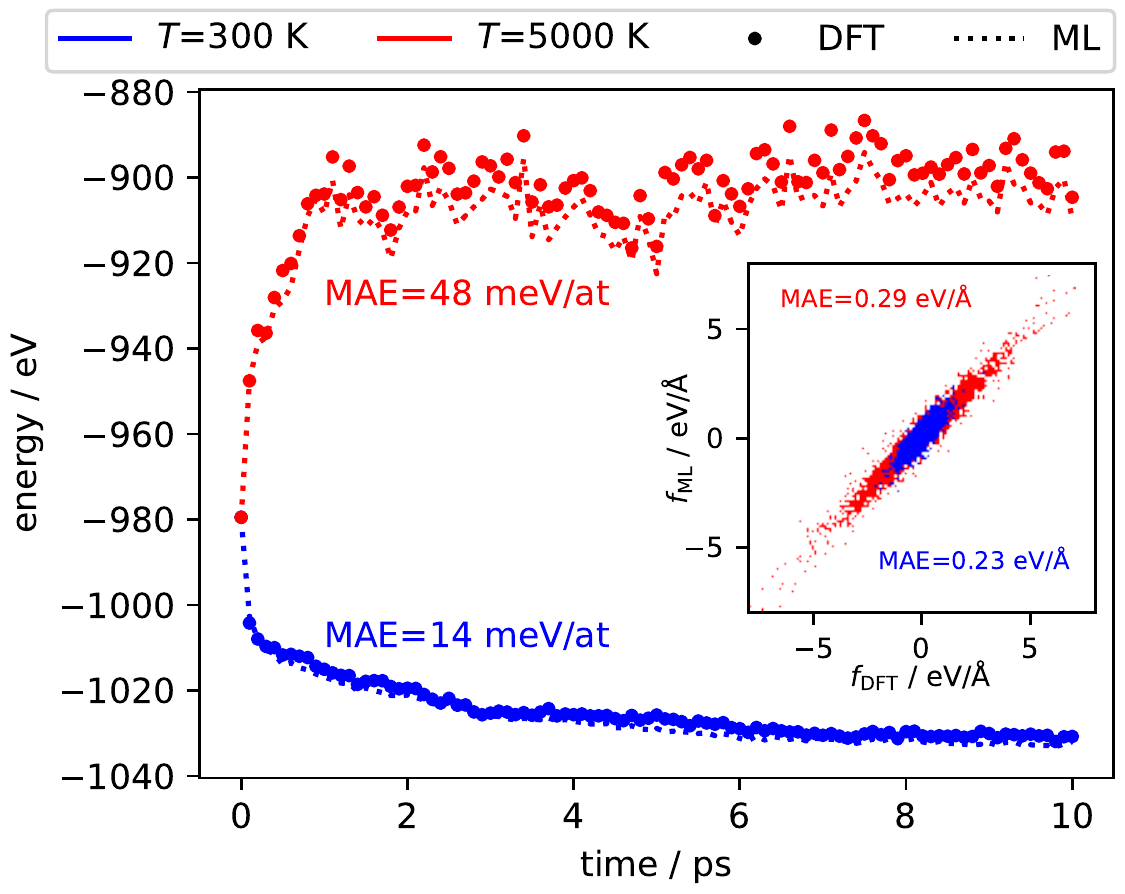}
    \caption{Comparison between the potential energy evaluated along two 10ps MD/MC trajectories, and that recomputed by DFT for 100 snapshots. The inset shows the parity plot for the force components computed for those structures. Energies have a MAE of 14 (48) meV/atom and forces a component MAE of 0.23 (0.29) eV/\AA{} for the 300 (5000) K trajectory. }
    \label{fig:validation-md}
\end{figure}

\subsection{Molecular dynamics}

As a further demonstration of the accuracy and the stability of this potential, we perform two constant-pressure MD/MC  trajectories, one at $T=300$K and one at $T=5000$K, each starting from a random arrangement of 5 atoms for each of the 25 elements (a total of 125 atoms) arranged on an \emph{fcc} lattice.
The trajectories are 10ps long, with on average one attempt at exchanging a pair of atoms every 2fs. We save a configuration every 100fs, and perform DFT calculations to compare energy and forces with those obtained from the ML potential.
Fig.~\ref{fig:validation-md} shows that the low-temperature trajectory, where major rearrangements of the atoms occur but the structure remains approximately \emph{fcc}, has an accuracy comparable to that measured on the validation set. The high-temperature run exhibits a higher error. However, the main component of the error is a rigid shift of the energies, and the trajectory remains stable -- which is remarkable given that we observe complete melting, and the potential is trained exclusively on distorted solid structures.

\begin{figure}[bp]
    \centering
    \includegraphics[width=1.0\columnwidth]{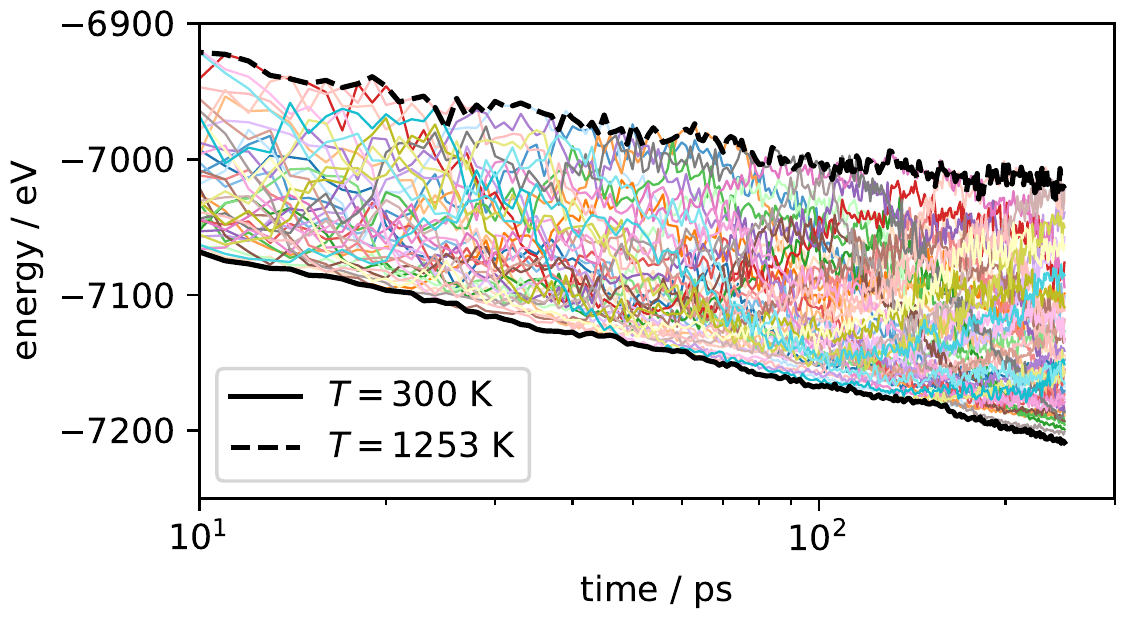}
\caption{
 Trajectories of the potential energy for the 40 replicas used in one of the REMD simulations of a 864-atoms box of the \heatf{}. 
 \rev{Each color corresponds to a different initial configuration, that goes through cycles of heating and cooling due to REMD exchanges, accelerating the equilibration of the simulation at each temperature.} The collection of trajectory segments corresponding to the extremal temperatures $T=300$~K and $T=1253$~K are highlighted with thicker, black lines. 
 The logarithmic time scale refers to the MD integration time, but should not be interpreted as physical time given the presence of MC steps and replica exchange moves.  }
    \label{fig:hea25-remd}
\end{figure}

\section{Temperature-dependent segregation in a Cantor-style alloy}
\label{sec:cantor}

In a seminal experiment, Cantor et al.\cite{cant+04msea} investigated the development of microstructure during the solidification of equimolar mixtures of 16 and 20 elements. 
We aim to perform a similar experiment in a computational setting, assessing the propensity of different elements to pair together or segregate, while covering the full component palette allowed by our model. 
This poses considerable challenges beyond the chemical complexity: kinetic trapping plays an important role in the physics of HEAs, and simulating vacancy-assisted atom diffusion requires time scales that are unattainable in brute-force atomistic modeling. 
In order to accelerate sampling and achieve (partial) equilibration, we run replica exchange simulations combining molecular dynamics and atom swap moves (REMD/MC), as described in Section~\ref{sub:sampling}.

Fig.~\ref{fig:hea25-remd} shows a representative trajectory for a 864-atoms cell, starting from \emph{fcc} configurations, and including equimolar composition of all 25 elements (a composition we will refer to as \heatf{}). 
The slow, logarithmic relaxation of the low-temperature replica is indicative of the glassy dynamics of the system, which does not equilibrate completely even after millions of MD/MC steps (see the \SM{}). %
For this reason, we perform multiple independent (and longer) simulations with a smaller box size \mc{(see the \SM{}).} The qualitative observations on the local ordering are robust, even though the precise arrangement of atoms in the low-temperature regime, as measured by the element-resolved pair correlation functions, differ noticeably between trajectories. 

\begin{figure}[tbhp]
    \centering
    \includegraphics[width=1.0\columnwidth]{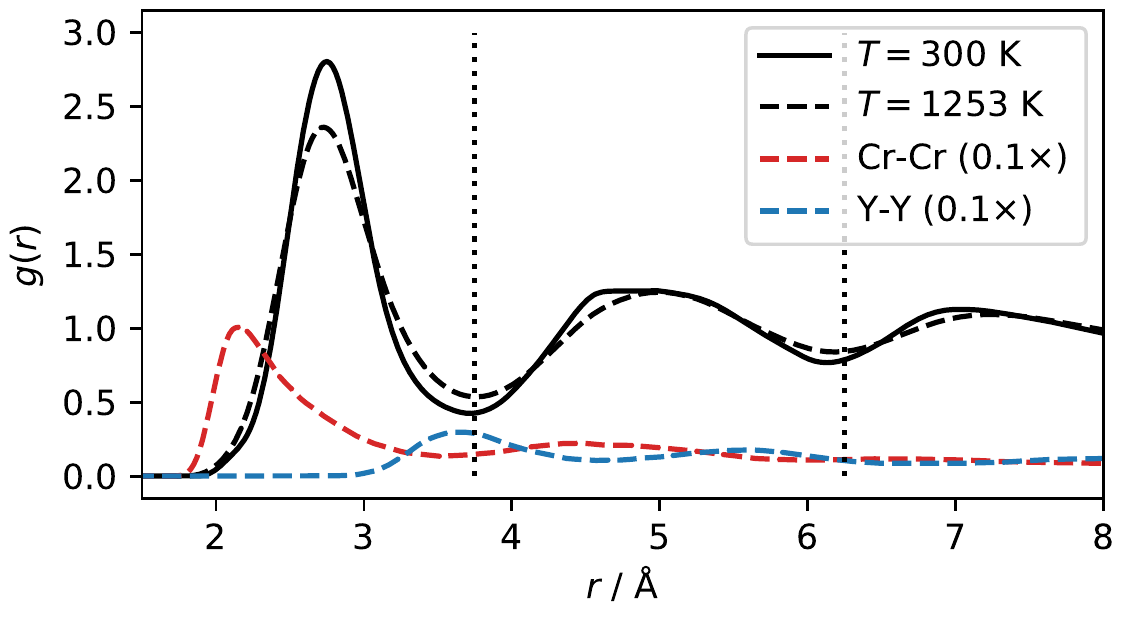}
    \caption{Pair correlation functions computed on a the $T=300$~K (full) and $T=1253$~K (dashed lines) replicas of a \heatf{} box. Black lines correspond to the unresolved pair correlation, while red (Cr-Cr) and blue (Y-Y) lines provide representative examples of pair correlations resolved by species.  The vertical dotted lines indicate the regions used in the definition of the pair ordering.  }
    \label{fig:hea25-gofr}
\end{figure}    

\subsection{Relative pair probabilities for the \heatf alloy}

The pair correlation functions (Fig.~\ref{fig:hea25-gofr}) display broad, liquid-like peaks at both the highest and the lowest temperature we considered. In fact, simulations show little diffusion (except for some occasional bursts of activity at the high end of the temperature range) and the system can be characterized as an amorphous (or nano-crystalline) solid. 
The broadening of the peaks can be at least in part attributed to the diversity of pair distances between atomic species: some, like Cr-Cr, peak at distances as short as 2\AA, others, such as Y-Y, peak at about 3.7\AA{}. Note that typical distances in same-element pairs do not always match those found in the pure solid, underscoring the fact that the \heapot{} can capture the effects arising from the heterogeneous chemical environments found in this alloy. 
For this reason, and given the disordered structure that develops in the supercell, we analyze structural correlations using a coarse-grained definition in which the first coordination shell extends up to a distance $r=3.75$~\AA{}, the second up to $r=6.25$~\AA{} and the third up to $r=8$~\AA{}, which is the largest distance we  consider given the size of the box.  
\newcommand{\DR}{{\Delta r}}
We then define a variation on a theme of the short-range order parameter~\cite{Cowley1950}, which we dub the relative pair probability (RPP) 
\begin{equation}
\text{RPP}_\DR(\ce{A},\ce{B}) =  \frac{p_\DR(\text{A},\text{B})}{p_\DR(\star,\star)}
\frac{\rho^2}{\rho_{\text{A}}\rho_{\text{B}}}
\end{equation}
which computes the number of pairs between species \ce{A} and \ce{B} that occur within a range $\DR$ of distances, divided by the number of all pairs found in that same region, and normalized by the number density of the two species, $\rho_{\text{A,B}}$ and the overall number density $\rho$.
$\text{RPP}=1$ indicates that the two species are as likely to be found within a given separation range than any atom pair. $\text{RPP}>1$ ($<1$) indicate that they are more (less) likely to be found in that distance range. 

\begin{figure}[tbhp]
    \centering    \includegraphics[width=1.0\columnwidth]{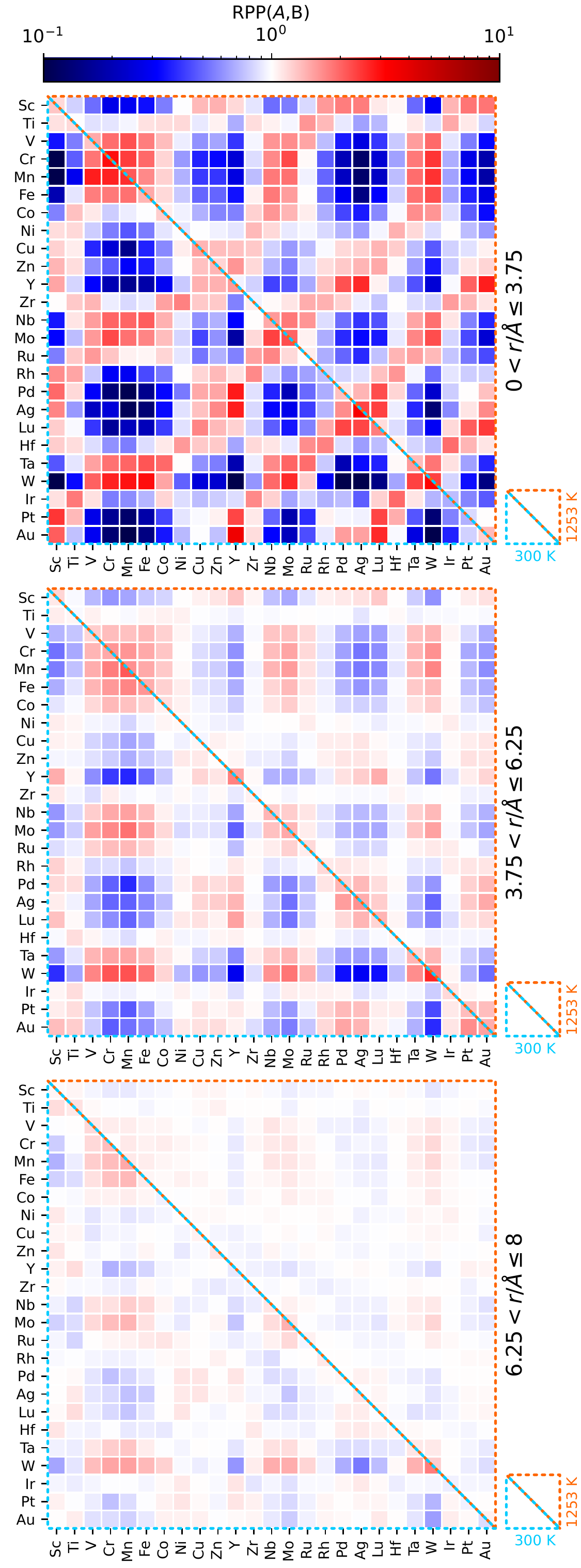}
    \caption{
    A plot of the relative pair probability for all atom pairs and the three regions corresponding to the first, second, and third peaks in the total pair correlation function ( Fig.~\ref{fig:hea25-gofr}). Each plot is split across the diagonal, showing results for simulations of \heatf{} at two temperatures, 300~K (lower-left corner) and 1253~K (top-right corner), averaged over the trajectories and discarding the first 100 ps (50'000 combined MD/MC steps).   }
    \label{fig:hea25-order}
\end{figure}

Qualitatively, the value of the RPP in the first coordination shell is indicative of the propensity of two elements to cluster together or to separate from each other. However, the values cannot be interpreted in isolation, without considering the overall setup of the simulation: the finite size of the supercell, the imperfect equilibration, and the many-body interactions between all 25 species mean that the strong affinity between Y and Au, or the poor compatibility of Mn and Pd, do not necessarily imply the same quantitative effect when considered as part of a different overall composition. 
Fig.~\ref{fig:hea25-order} shows a heat-map representation of $\text{RPP}_\DR(\ce{A},\ce{B})$ for the \heatf{} at 300~K and 1253~K, and for the three regions indicated in Fig.~\ref{fig:hea25-gofr}. 
A few qualitative observations can be made. First, in our simulations \heatf{} evolves to be far from random. Certain atom pairs have a strong tendency to associate or separate at low temperature, and the high-temperature samples (which are well equilibrated) show similar, even though less pronounced, trends. 
This correspondence is interesting, as it suggests one may use high-temperature trajectories, that are easier to converge, to extract insights on the propensity of different species for association. 
The trends observed in the second and third region are very similar to those in the first-extended-neighbor shell, although progressively less pronounced: given the finite size of the simulation, and incomplete equilibration, the simulation does not generate clear-cut phase-separated regions.

Considering the RPP along the elements, one can observe a clear periodicity in behavior. Sc, Y, Hf, as well as the noble metals, Cu and Zn, tend to separate from V, Cr, Mn, Fe, which on the other hand have a tendency to cluster together, and also have positive associations to their heavier counterparts Nb, Mo, Ta, W. 
On the other hand, Sc, Y and (to a lesser degree) Hf associate strongly with noble metals, Cu, and Zn. The noble metals, Cu and Zn also tend to cluster together.
Ti, Co, Ni, Zr, Ru, Ir have less clear-cut associations, and are closer to having a random distribution throughout the box.  
Another way of looking at the association plots in Fig.~\ref{fig:hea25-order} is to check for consistency with known high-entropy alloys. 
The Cr-Mn-Fe-Co-Ni system is one of the prototypical sets of HEA formers, and indeed we observe strong mutual association tendency between Cr-Mn-Fe in the first shell, and also with Co and Ni in the second extended shell. 
Second-shell mutual association is also observed for noble-metal based compositions such as Ni-Cu-Pd-Pt-Au.
Let us reiterate that strong mutual association for a group of elements in the \heatf{} runs is a necessary, but not sufficient, conditions for that group of elements to be good HEA-forming candidates. For instance, some elements may have a strong tendency to form ordered intermetallics and might separate out of the mixture.

\begin{figure}[b]
    \centering
    \includegraphics[width=1.0\columnwidth]{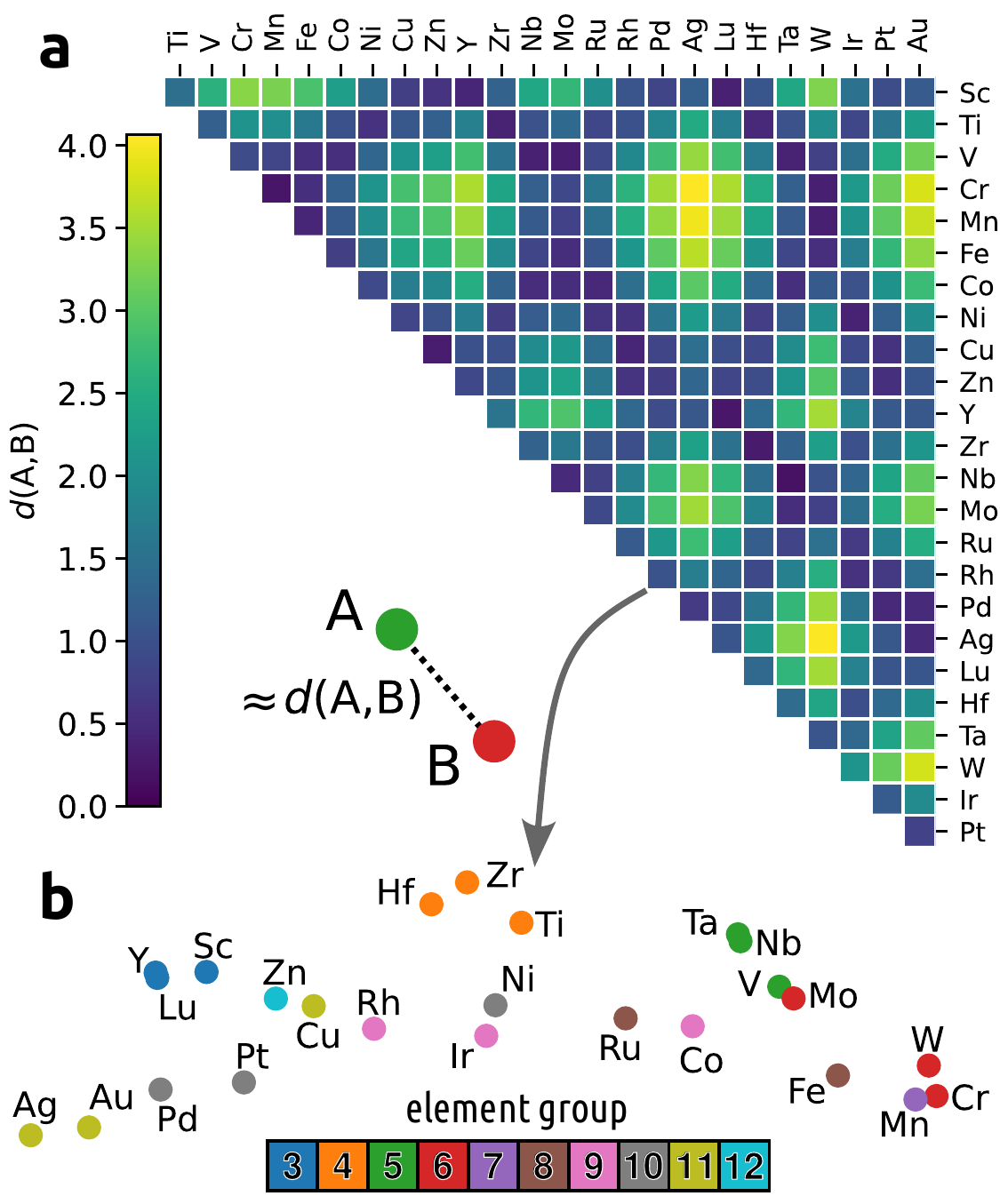}
    \caption{ (a) Element similarity matrix based on the RPP distance~\eqref{eq:d-rpp} for the nearest-neighbor shell, in the \heatf{} simulation at $T=1253$~K. (b) The element similarity map (color-coded based on the group of the various transition metals) is built by applying metric multi-dimensional scaling to the distance matrix, and provides a visual aid to recognize groups of elements that have similar affinity patterns to the other $d$-block metals. 
       }
    \label{fig:hea25-hr}
\end{figure}

\subsection{Data-driven Hume-Rothery rules}

This analysis allows us to substantiate and quantify some of the empirical principles that are used in the design of HEAs, such as Hume-Rothery rules\cite{Mizutani2012} that stipulate what elements can be substituted for each other with little effect on the HEA-forming propensity.  
We use the first-neighbor affinity of each species to all the other elements in the alloy to define a measure of dissimilarity as 
\begin{equation}
d_\text{RPP}(\text{A,B})^2 = {\sum_\text{X} \left[\log_{10}\frac{\text{RPP}_1(\text{A,X})}{\text{RPP}_1(\text{B,X})} \right]^2  },
\label{eq:d-rpp}
\end{equation}
that, roughly speaking, measures the relative strength of interactions between the two species and the other components. 
Two elements with a small distance are predicted to behave similarly, and vice versa. Fig.~\ref{fig:hea25-hr} paints a picture that is consistent with the observations we made on short and mid-range order between the elements in the \heatf{}, and with much of the common wisdom in HEA research.%
We base this analysis on the high-temperature simulations to obtain a statistically-converged, and somewhat more nuanced, definition, but the qualitative features of the map are similar to those one would obtain from the RPP computed at $T=300$~K.
Elements in the same group usually show strong similarity, but this is not always the case: for example, Cu is more similar to Zn than to Ag. 
The similarity matrix can also be converted to a 2D  map, in which the Euclidean distance between elements approximates their RPP-based similarity (also shown in Fig.~\ref{fig:hea25-hr}), which provides an easy-to interpret visual representation of a set of data-driven rules to design HEAs. 
The element similarity that can be inferred from the RPP-based map differ -- both quantitatively and conceptually -- from that associated with the alchemical coupling matrix in Fig.~\ref{fig:coupling-weights}. Whereas the weights are associated with the similarity in terms of the interatomic potential, the RPP similarity is a result of the collective behavior of the \heatf at the prescribed thermodynamic conditions, not unlike the relation between a pair potential and the potential of mean force. This means, for example, that one could compute $d_\text{RPP}$ for a different alloy composition (extending or refining the assessment of alloying behavior), from a different type of interatomic potential, or even from experimental data on partial structure factors.

\section{Bulk structure of high-entropy alloys for catalysis}
\label{sec:app}

Having demonstrated the accuracy of the \heapot{} model, and used it to investigate the mutual affinity of the full set of 25 transition metals we considered in a Cantor-type computational experiment, we now turn our attention to a more focused study of three specific equimolar compositions. 
The first is the prototypical CoCrFeMnNi alloy, which was reported by Cantor et al.\cite{Cantor2004} in their seminal paper. This alloy is also known to be effective as a catalyst\cite{Waag2019,Peng2020,He2022}. Furthermore, we investigate CoCrFeMoNi\cite{Zhang2018,Tang2021,Schumacher2022},as an example of an alloy obtained by element substitution that has been broadly studied for its improved mechanical and tribological properties \cite{Cui2020,Wang2022}, as well as a catalyst of oxygen evolution reactions. 
We then consider IrPdPtRhRu\cite{Batchelor2019,Broge2020,Wu2020,Pedersen2021,Lee2022} as an example of an alloy based on sixth period elements that has recently received much attention as a catalyst for hydrogen evolution, and is often synthesized in the form of nanoparticles.

\begin{figure}[tbhp]
    \centering
    \includegraphics[width=1.0\columnwidth]{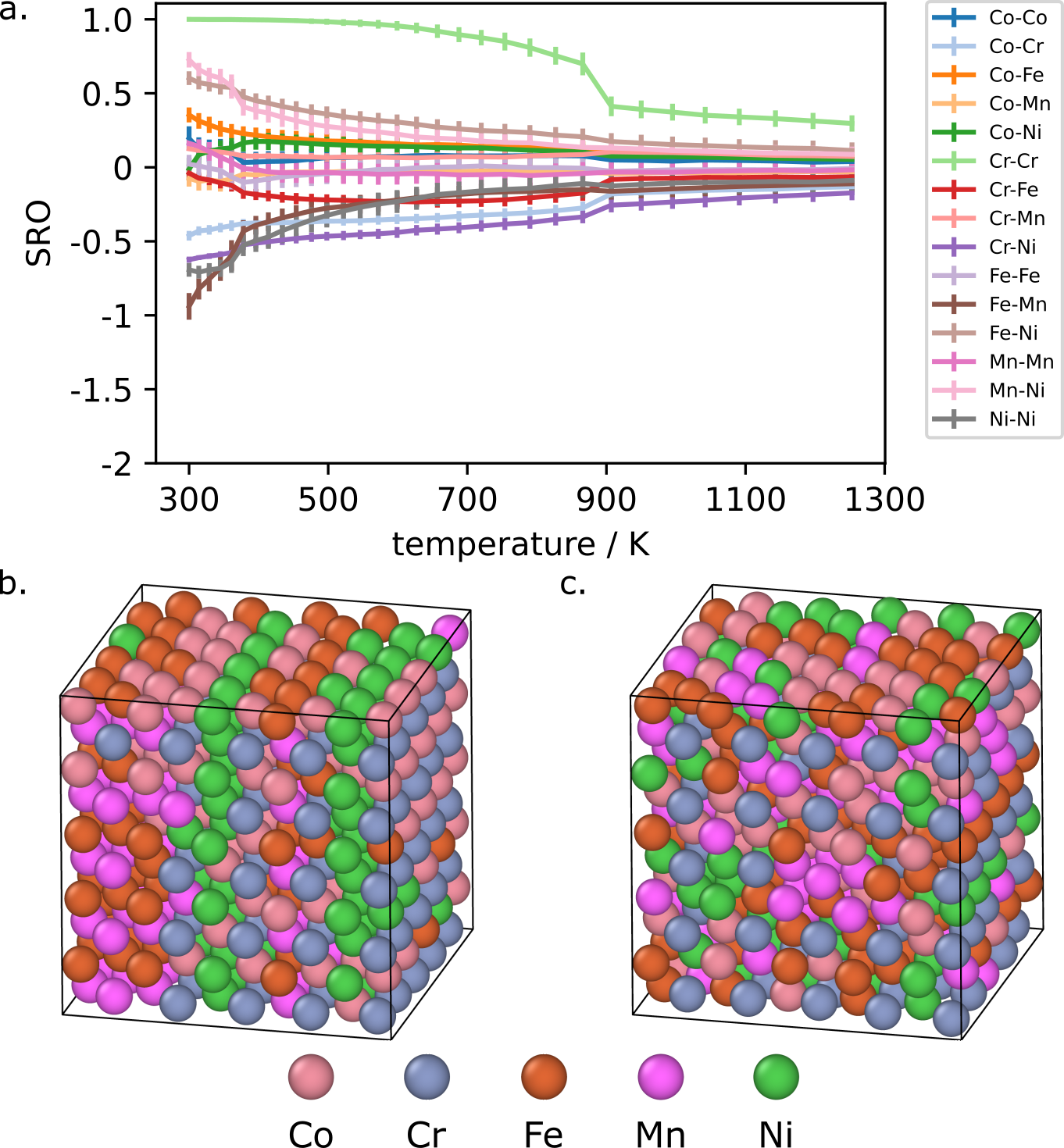}
    \caption{
    a. Cowley's short-range (SRO) parameters for the first shell in CoCrFeMnNi HEA, shown for the 10 replicas between 300 and 1253~K, averaged over the last 1000 steps and two independent runs. At low temperatures, a tendency of Fe-Mn segregation can be seen. In contrast, Co is very well mixed. There are two phase transformations around 400~K and 900~K. The y-axis is adjusted to the example shown in Fig.~\ref{fig:irpdptrhru-sro} to facilitate comparison. b,c. snapshot from MC/MD simulations at $T=300\,\si{\kelvin}$ and at $T=720\,\si{\kelvin}$, respectively. In the 300~K snapshot, two planes of Ni can be seen, while in the higher temperature snapshot, Cr order is evident (see the \SM{}).    
    }
    \label{fig:cocrfemnni-sro}
\end{figure}

To model the alloys, we used \textit{fcc} lattices with 500 atoms per cell ($5 \times 5 \times 5$ super cell). We ran two independent REMD/MC runs according to Section~\ref{sub:sampling} with a timestep of 2~fs and 32 temperature replicas, logarithmically spaced between 300~K and 1253~K. 
We discard the first 100ps for equilibration. 
Given that all these alloys maintain a regular \emph{fcc} structure throughout the simulation, we analyze their structure in terms of  Cowley's short-range order\cite{Cowley1950} (SRO) \rev{which is commonly used in the study of HEAs and takes a value of zero when atoms are distributed fully randomly, becomes negative for pairs of atoms that tend to cluster together, and tends to one when two atom types never appear as first neighbors}. 
In the \SM{} we also report an analysis in terms of the RPP that incorporates second-neighbor and long-range correlations. 
In interpreting these results, one should consider similar considerations to those we discussed for the \heatf{} simulations: (1) the SRO (and the RPP) are only meaningful for homogeneous phases, and in case of phase separation the values computed for the whole cell serve only to signal the occurrence of a phase transition; (2) a combination of finite-size effects and glassy behavior can hinder reaching full equilibrium in simulations; (3) since they allow for atom exchanges, our simulations cannot give quantitative indications on whether different phases are only metastable, nor on the kinetics of diffusion processes that are required for precipitation.

We start by analyzing the Cantor alloy CoCrFeMnNi. 
The SRO computed at different temperatures (Fig.~\ref{fig:cocrfemnni-sro}a, plotted for all element combinations) indicate the presence of at least two phase transitions. The high-temperature phase is homogeneous and disordered, but shows substantial ordering, particularly for the Cr-Cr pair. 
At approximately 900~K we observe a first transition, that is associated with the ordering of Cr atoms. The SRO for the Cr-Cr pair tends to one (as there are almost no first-neighbor chromium atoms) but the RPP show a clear increase of second-neighbor Cr-Cr pairs, consistent with the formation of a simple cubic sublattice. The other elements remain relatively disordered, and no discontinuous behavior is observed in the SRO.  
As the temperature is reduced further, a second transition occurs around 400~K. The most prominent structural transformation is the formation of (100) Ni planes, separated by (Co,Fe,Mn)-rich regions forming a layered superstructure. 
Fig.~\ref{fig:cocrfemnni-sro}b,c show snapshots of the simulations at 300~K  and 720~K, that give an idea of the partially-ordered structure of the two phases.

\begin{figure}[tbhp]
    \centering
    \includegraphics[width=1.0\columnwidth]{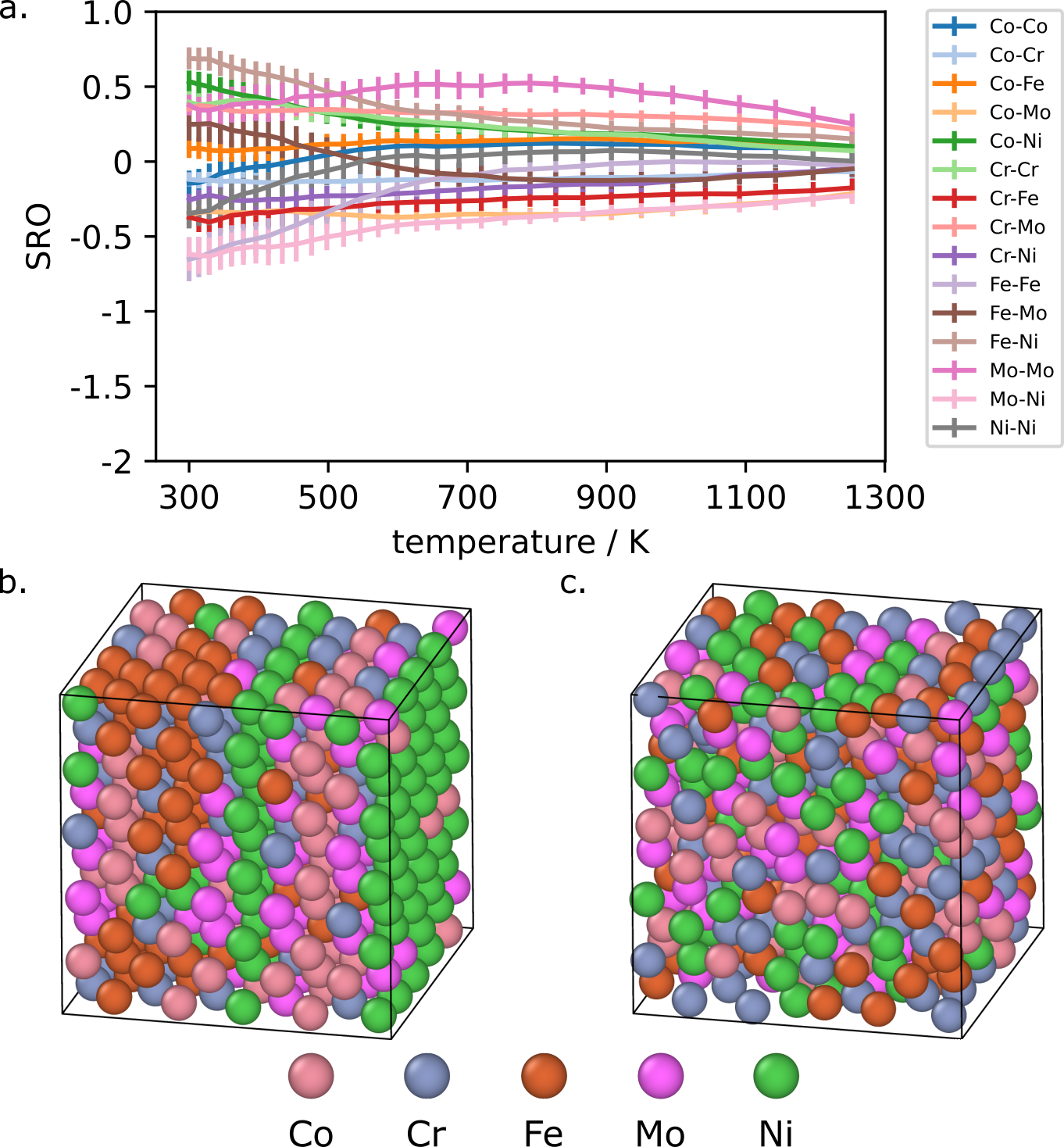}
    \caption{
    a. Cowley's short-range (SRO) parameters for the first shell in CoCrFeMoNi HEA, shown for the 10 replicas between 300 and 1253~K, averaged over the last 1000 steps and two independent runs. Good mixing of atomic species can be assumed due to the small values of SRO parameters. The y-axis is adjusted to the example shown in Fig.~\ref{fig:irpdptrhru-sro} to facilitate comparison. b,c. snapshot from MC/MD simulations at $T=300\,\si{\kelvin}$ and at $T=1253\,\si{\kelvin}$, respectively. In the 300~K snapshot, two planes of Ni can be seen.}
    \label{fig:cocrfemoni-sro}
\end{figure}

Substituting Mn with Mo changes the segregation behavior significantly (Fig.~\ref{fig:cocrfemoni-sro}a): the SRO parameters are generally smaller, with the largest segregation tendency found for the Mo-Ni atom pair. 
The tendency of Cr to form a cubic sublattice is less pronounced than CoCrFeMnNi, and one only sees the increase of SRO parameters at around 500~K. At low temperature, (100) planes of Ni form that are very similar to those observed in the Mn-based counterpart (Fig.~\ref{fig:cocrfemoni-sro}b,c), that are separated by  (Co,Fe,Mo)-rich regions.
Given the sizable energy errors of the ML models, as well as those of the underlying DFT reference, one should not overinterpret the details of the structures we observe. 
\rev{Even if \emph{fcc} CoCrFeMnNi is paramagnetic, neglect of magnetism in the presence of several elements which form ferromagnetic phases is worrisome (see e.g. Ref.~\citenum{sebe+19prm} for a thorough discussion of magnetism in CoCrFeMnNi and CoCrFeMoNi).
That said, our observations provide strong indications of the tendency to form partly ordered phases with a complex structure, which, together with the low vacancy-mediated diffusivity,\cite{tsai+13am} help explain the observed stability of HEAs that contain (Co,Fe,Cr,Ni). A tendency to develop short-range ordering is consistent with previous simulations in other classes of HEAs,\cite{chen+21ncomm} and with observation of phase separation in equimolar CoCrFeMnNi in high-mobility environments such as grain boundaries\cite{glie+20am} or under deformation\cite{lee+19m}.
}

\begin{figure}[tbhp]
    \centering
    \includegraphics[width=1.0\columnwidth]{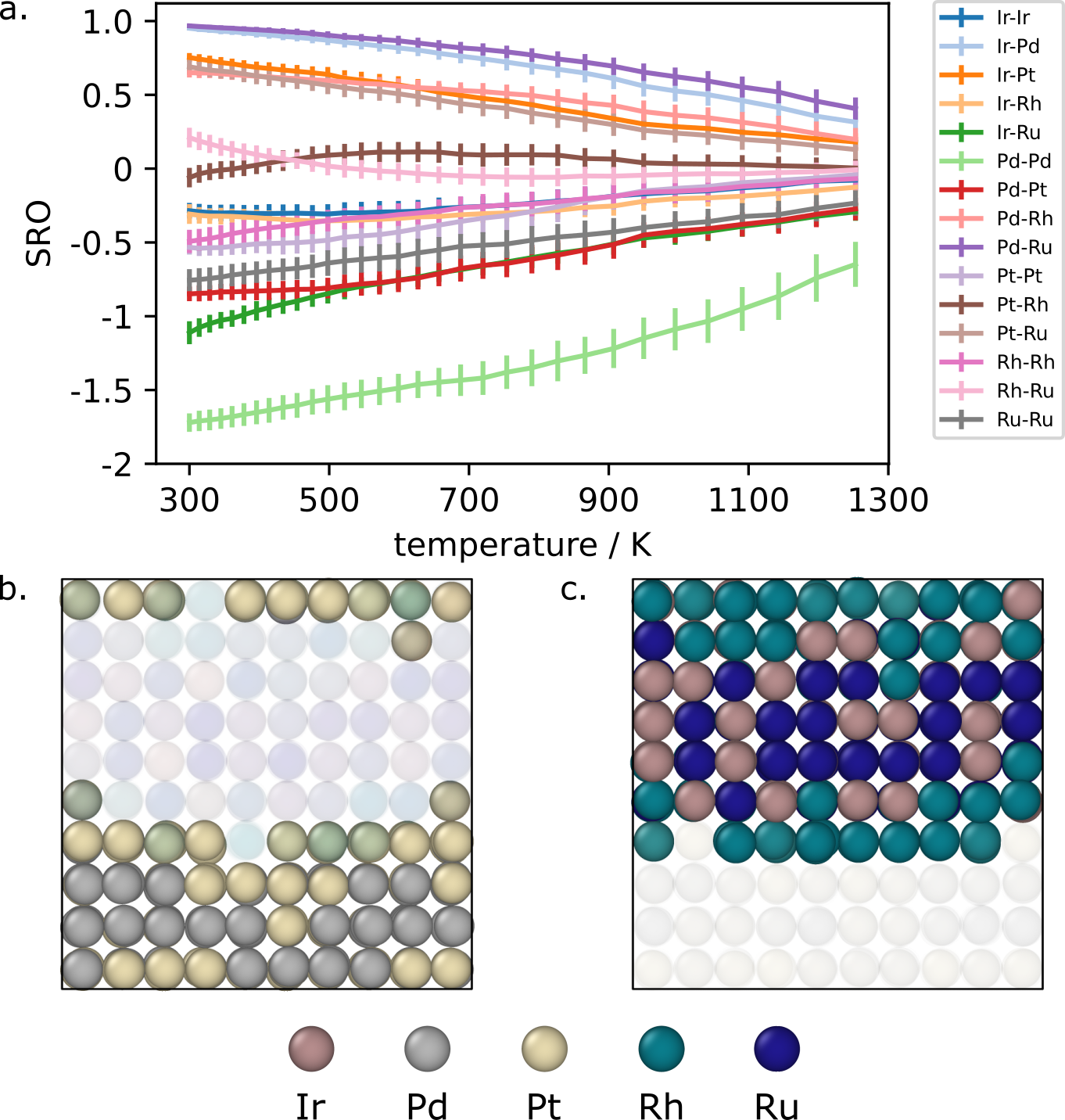}
    \caption{
    a. Cowley's short-range parameters for the first shell in IrPdPtRhRu HEA, shown for the 10 replicas between 500 and 933~K, averaged over the last 1000 frames and with an error estimation from independent repetition runs. The most pronounced local order can be seen for the Pd-Pd atom pair (light green line, mathematically smallest SRO). Demonstration of the phase segregation tendency by highlighting the b. PdPt and c. IrRhRu atoms in an MC/MD snapshot.}
    \label{fig:irpdptrhru-sro}
\end{figure}

While the leading effect in CoCrFeMnNi and CoCrFeMoNi is the appearance of partial ordering at low temperatures, in the case of IrPdPtRhRu we observe clear-cut phase separation betwen a (Pd,Pt) and a (Ru,Ir,Rh) phase, with Rh accumulating preferentially at the interface between the two phases (see Fig.~\ref{fig:irpdptrhru-sro}b,c).
The strong tendency to segregate is already evident in the high-temperature regime, where the system is visually well-mixed, but with large SRO parameters. 
This is in contrast to the experimental observation that this HEA forms a complex solid solution with random atom distribution\cite{Wu2020,Pedersen2021}. 
As shown in the \SM{}, the large enthalpic gain arising from demixing is not an artefact of \heapot{}, and the ML error on the free-energy change upon ordering is of the order of 3 meV/atom. 
These observations suggest that kinetic trapping, or finite-size effects associated with the synthesis in the form of nanoparticles, might be key to stabilize a homogeneous phase.

\section{Conclusions}

The notion that different chemical elements may behave similarly when combined with others is one of the founding principles of chemistry, and is often used as guidance in the design of new materials. 
We build a ML framework that incorporates this notion in the form of a linear compression of chemical space, and succeed in training a potential that can describe with semi-quantitative accuracy bulk phases of arbitrary combinations of 25 $d$-block elements. 
The physically-motivated, intuitive functional form of the contraction allows us to analyze critically the model performance, allowing us to show that 3-4 dimensions suffice to capture the diversity of behavior of the transition metal block. The optimized values of the combination weights reveal relationships between the elements that match their arrangement in the periodic table, to the point where we show it is possible to ``fill in the blanks'' for missing elements, with only a moderate loss of accuracy. 

We use the potential to run an ambitious computational experiment, in which we attempt to equilibrate an equimolar mixture of all 25 elements, resulting in the formation of a disordered structure with strong element segregation. 
The affinity between elements is consistent with several known high-entropy alloys, and allows us to define a data-driven version of the Hume-Rothery rules, that could be further adapted to subsets of elements that are relevant for a given application. 
We also investigate in detail three specific compositions - the archetypal Cantor alloy \ce{CoCrFeMnNi}, which we observe to be undergo a sequence of transitions towards complex ordered phases as the temperature is lowered; that arising from the Mn$\rightarrow$Mo substitution, which also leads to similar, although less pronounced, ordering; the noble metal alloy \ce{PdPtIrRuRh}, that shows a strong  tendency to decompose into into PdPt and IrRhRu phases.

We are only scratching the surface of what can be achieved within this framework. Extending the dataset to an even more diverse palette of compounds, and to structures that include molten and defective configurations, is an obvious direction for further improvements. A more systematic exploration of the design space of chemical compression is another promising research direction, even though doing so may sacrifice, at least in part, the interpretability of the linear contraction we use here. 
On a more application-focused front, a systematic study of the stability of 4 and 5-element HEAs along the same lines of the simulations of those which we present here, based on the current \heapot{} model, will provide much-needed insights into the stability range of multi-principal-component alloys, guiding synthetic efforts towards compositions that are stable towards phase separation. 

\section*{Data availability}

All data and code used to train the \heapot{} model, as well as the fitted parameters and code to run the simulations discussed in this work is available in the \SM{} or from publicly-accessible repositories (\url{https://github.com/Luthaf/alchemical-learning} and \url{https://archive.materialscloud.org/record/2023.57}).

\section*{Acknowledgements}

We thank William Curtin and Binglun Yin for sharing a dataset we used in the early developments of this work, and Anirudh Natarajan for discussion on an early version of the manuscript. MC and NL acknowledge support from the NCCR MARVEL, funded by the Swiss National Science Foundation (grant number 182892) and from an Industrial Grant from BASF. GF acknowledges support by the Swiss Platform for Advanced Scientific Computing (PASC). Electronic-structure calculations were performed within the scope of a CSCS project (ID: s1092).

\providecommand{\noopsort}[1]{}
\onecolumngrid\clearpage


\section*{Supplementary material}

### D. DFT validation of segregation in IrPdPtRuRh

In order to verify that the segregation we observe for IrPdPtRuRh is not an artifact due to the ML model, we ran similar simulations with a smaller supercell containing 180 atoms at 300 K. We performed a simulation with a simple MD trajectory, starting with a random distribution of atoms (random), and one including MC steps (swaps). After equilibration (which led to segregation in the simulation including atom swaps) we extracted 70 snapshots from both the "random" and "swaps" simulations, and re-computed their energies with the same DFT settings we used to generate the train set. It is clear that the energies are in very good agreement. There is a small near-constant shift in energies, but the relative energy of the two sets of configurations, as well as the spread within each subset, are reproduced with an accuracy comparable to that seen in the validation set. The large enthalpy gain associated with ordering is observed also in DFT calculations, which rules out the possibility that the observation of de-mixing is a consequence of errors in the ML model. We estimated the change in de-mixing free-energy due to the ML approximation using a free-energy perturbation expression, finding an error below 3meV/atom.

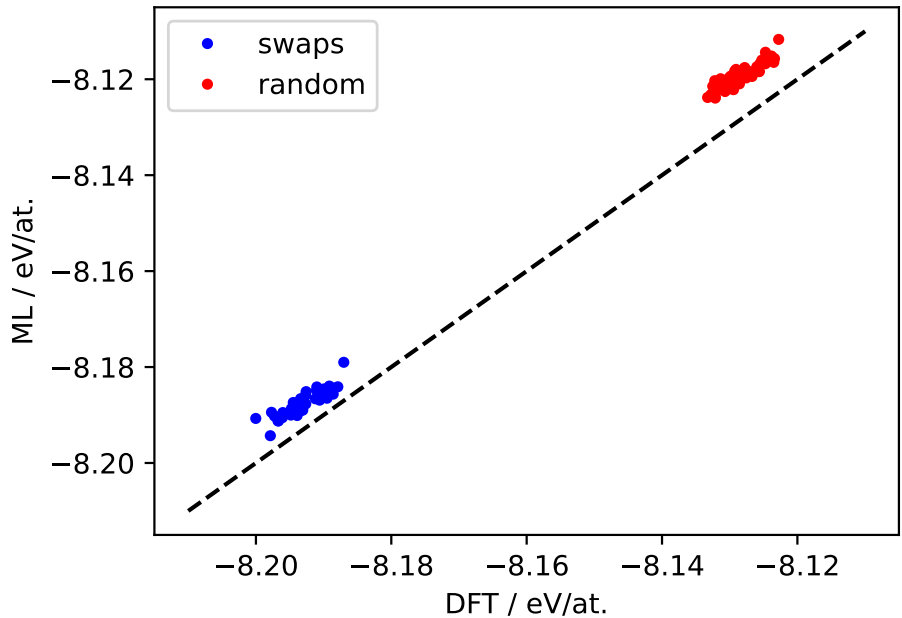

Figure S25. Parity plot of the HEA25-4-NN vs DFT energies for snapshots of a 180-atoms supercell containing IrPdPtRuRh, simulated at 300 K with plain MD (random) and including MC steps (swaps). A plot of the relative pair probability for all atom pairs and the three regions corresponding to the first, second, and third peaks in the total pair correlation function (Fig. S23). Each plot shows results for simulations of IrPdPtRhRu at both 500 K (lower-left corner) and 933 K (top-right corner), averaged over the trajectories and dis- carding the first 100 ps.

## VI. FULL LIST OF HULL PLOTS

Below you can find a full list of convex hull plots, that extend to the full list of 25 elements the plots shown in Fig. S9. Solid lines highlight the hulls obtained from the single-point DFT calculations (blue) and the ML predictions (purple). The dashed line identifies the structures that are stable based on the energies available in the Materials Project database. Note that in many cases there is a qualitative discrepancy not only between the HEA4NN model and the Materials Project reference, but also between the Materials Project reference and the internally-consistent DFT reference. Black points reflect discarded structures based on the distance measure reflecting the similarity to BCC or FCC phases.

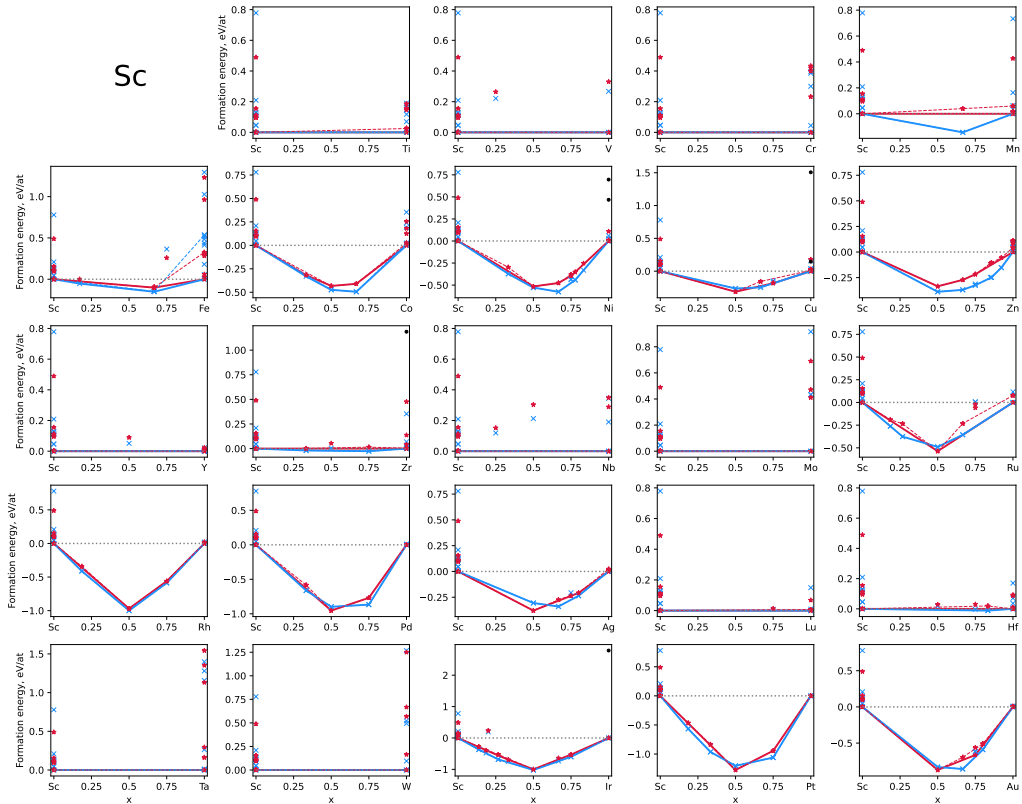

**Ti**

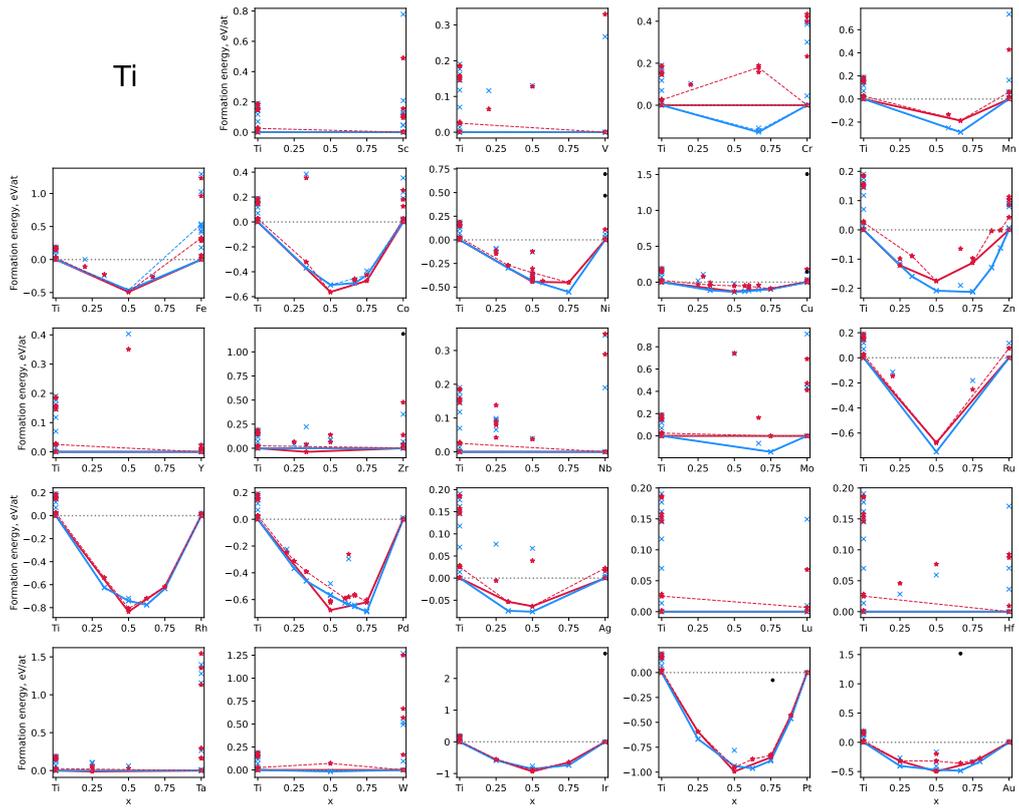

**V**

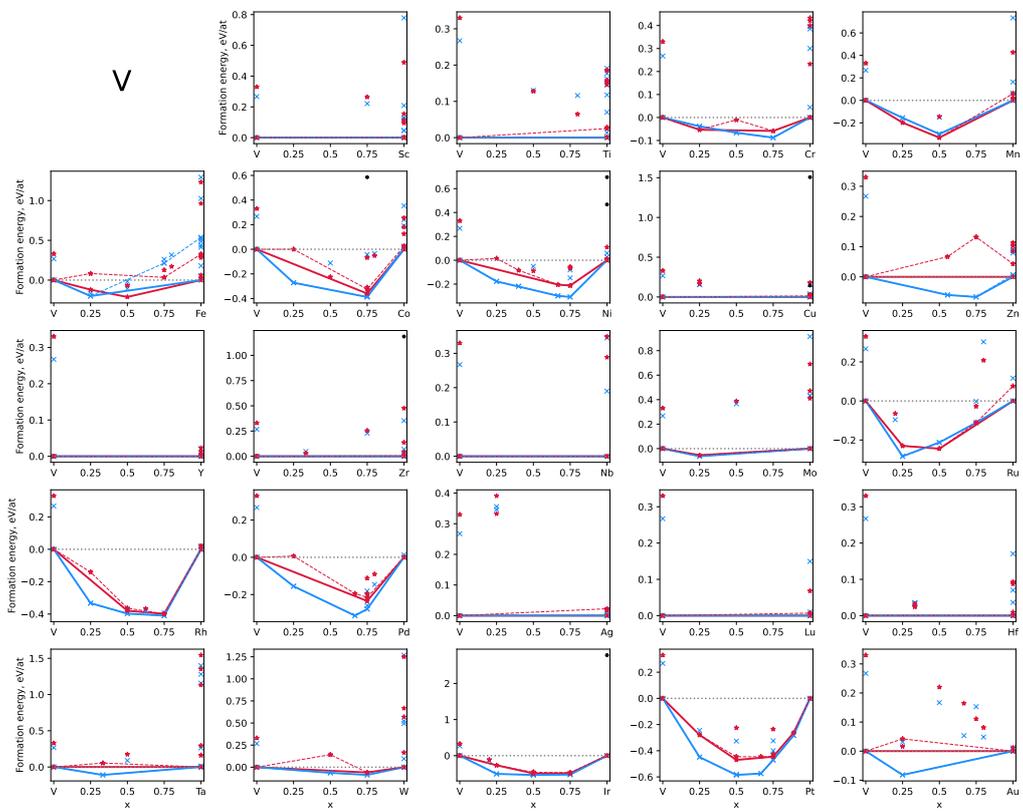

**Cr**

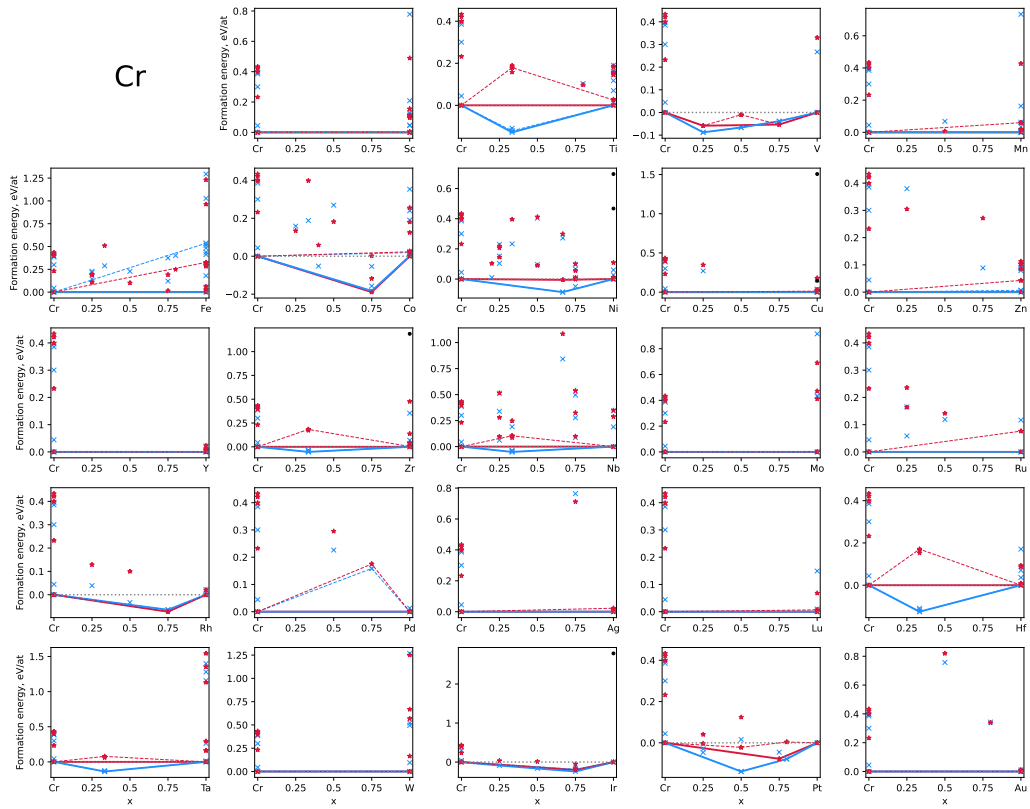

**Mn**

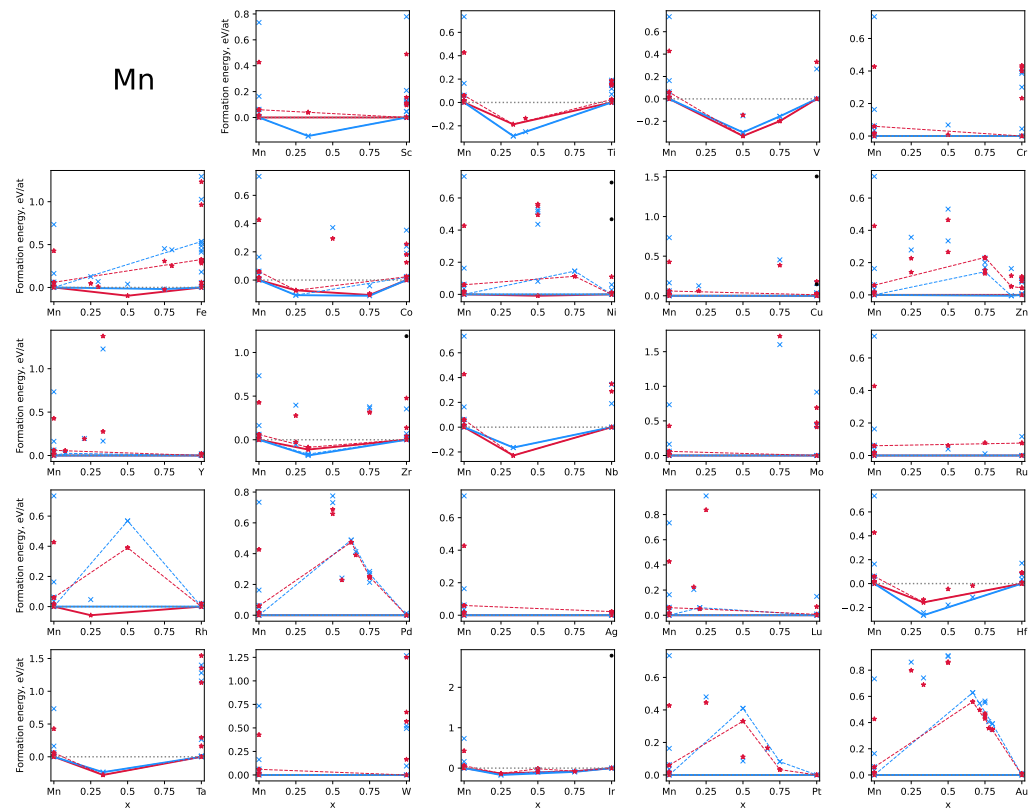

**Fe**

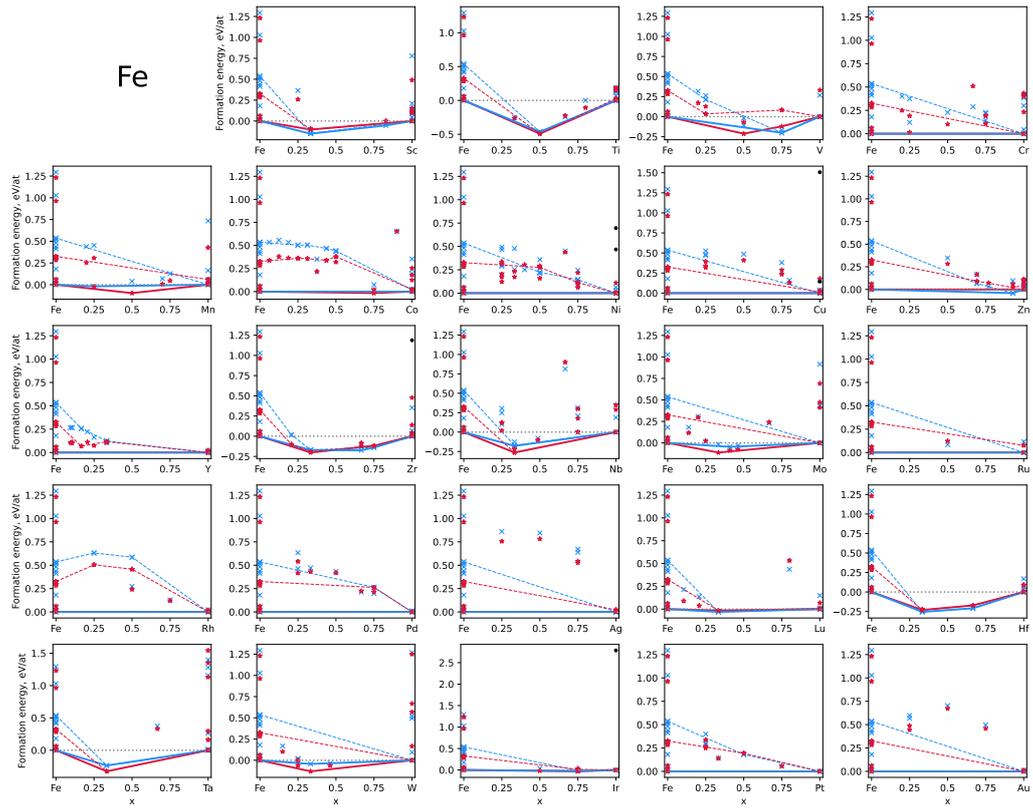

**Co**

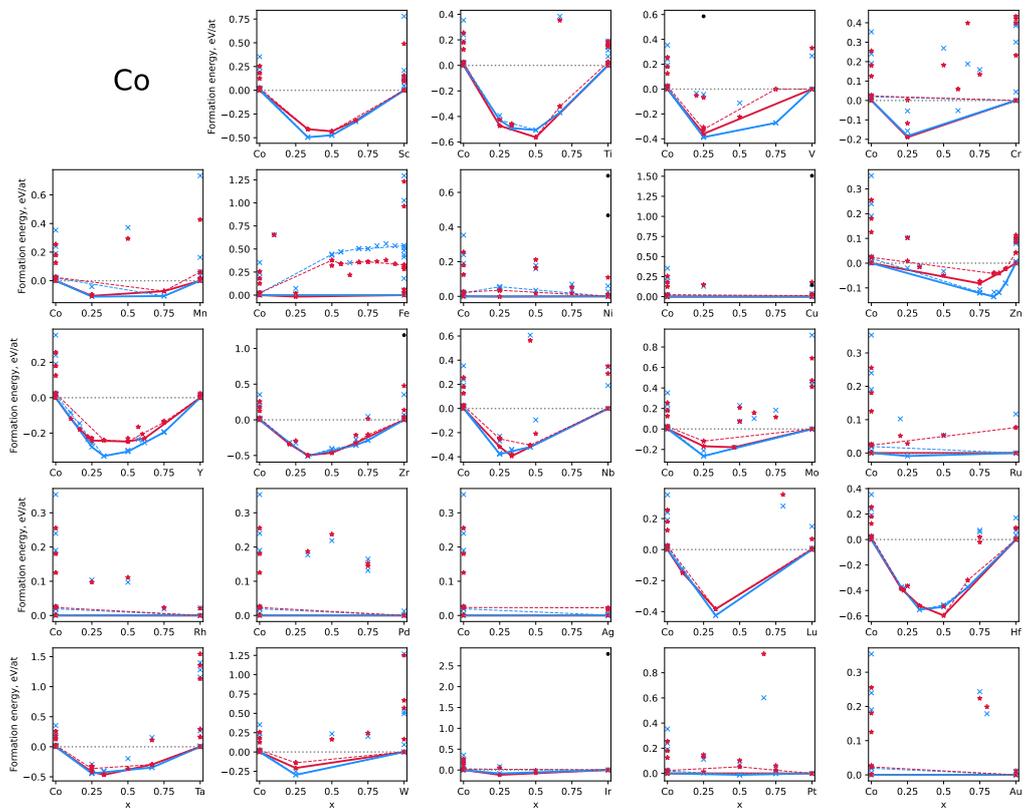





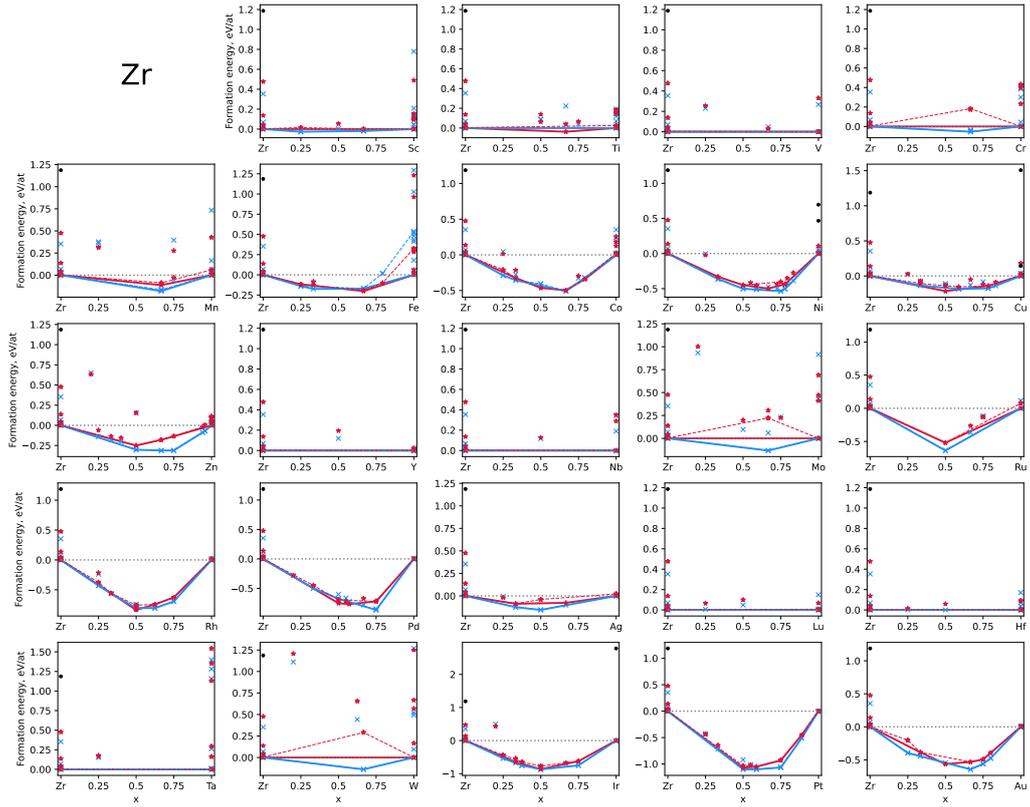

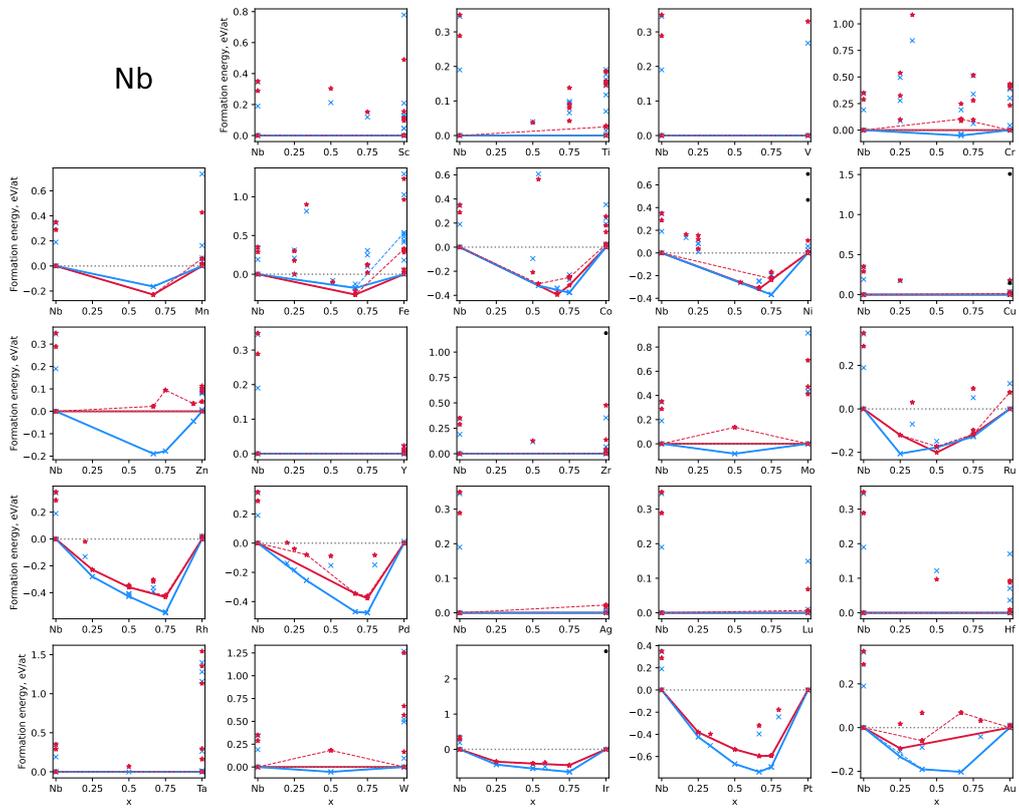

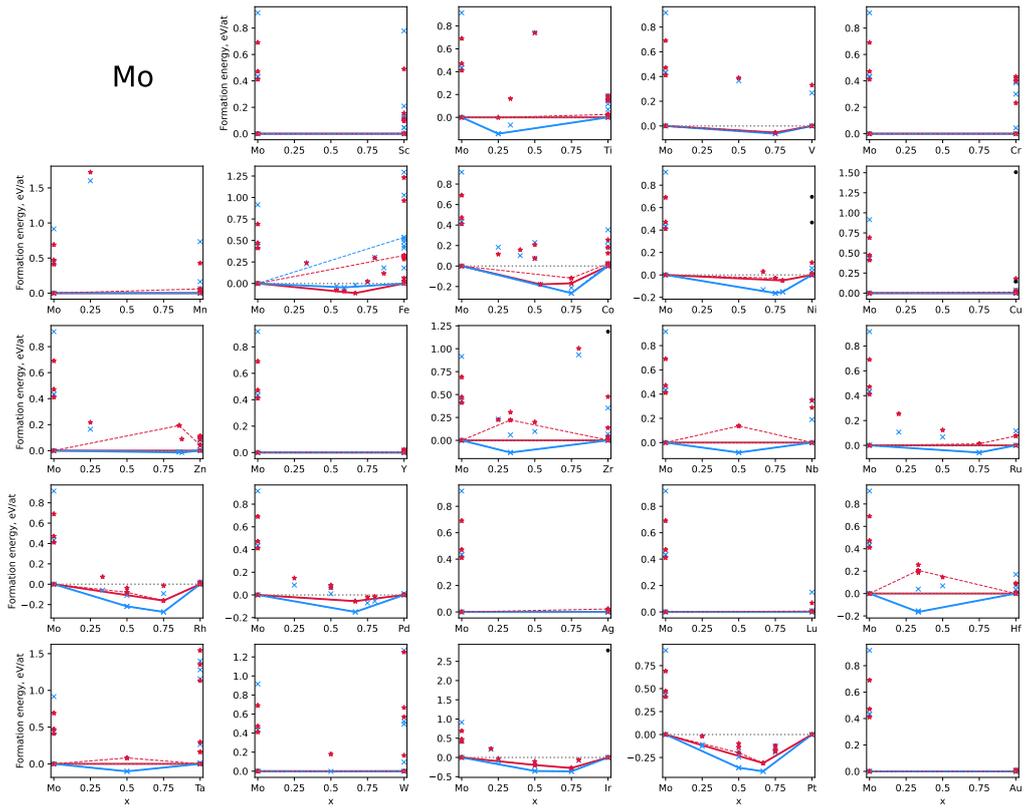

Mo

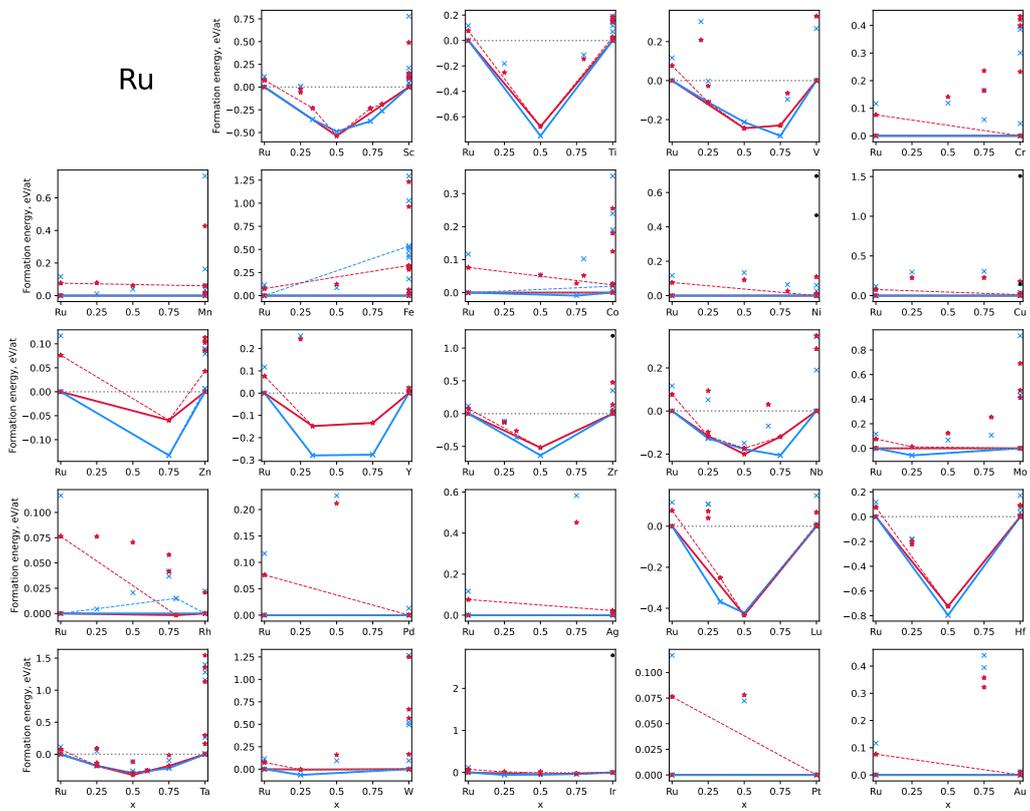

Ru

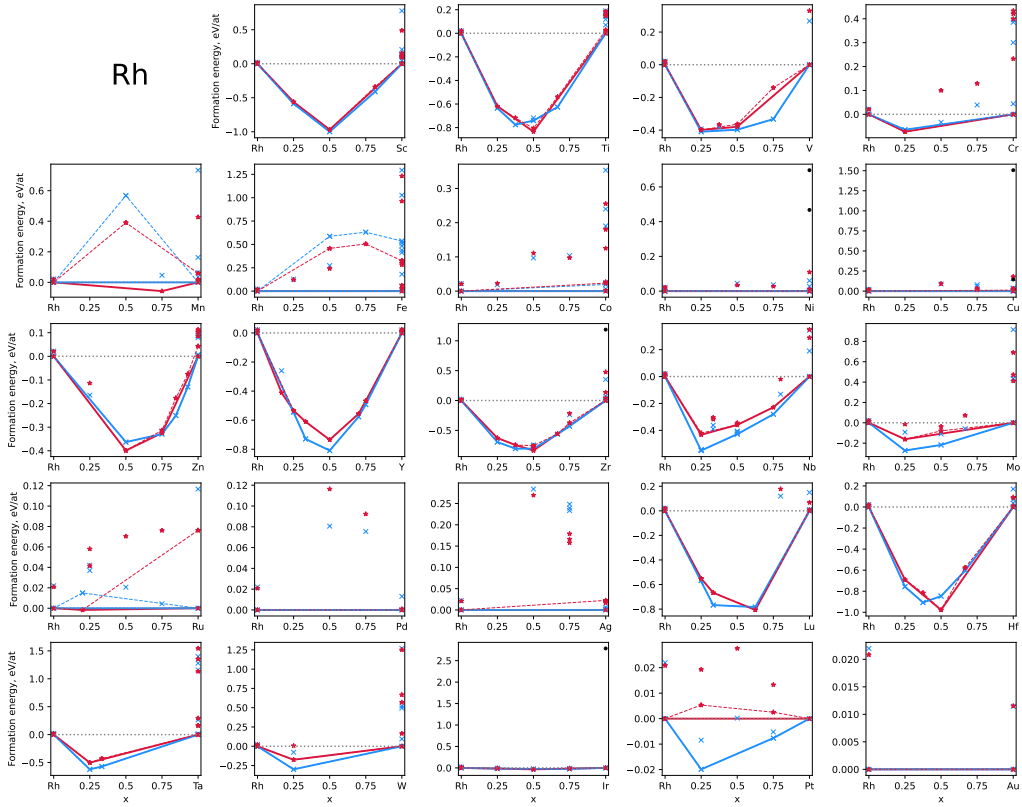

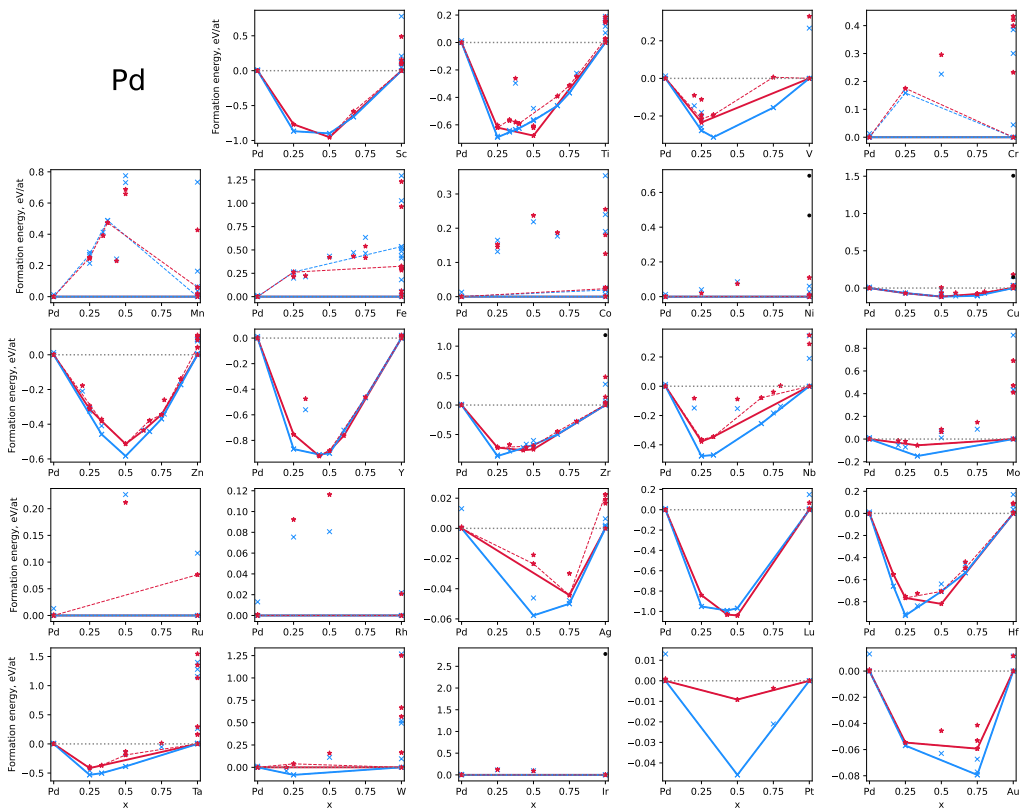

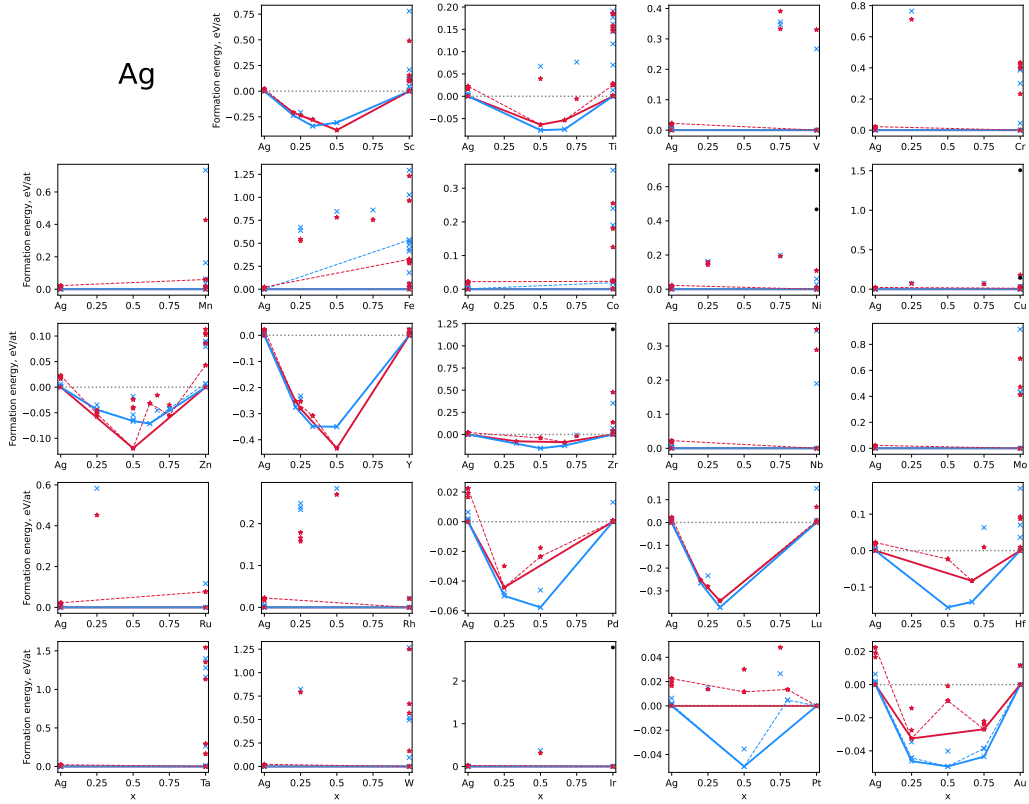

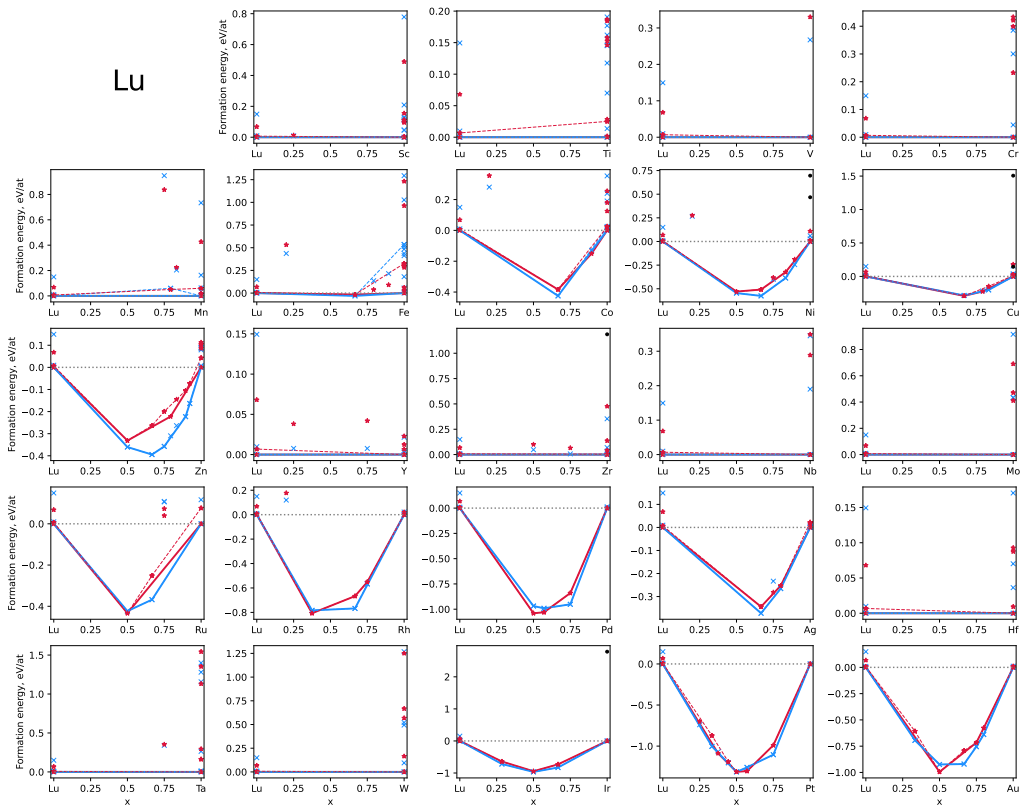

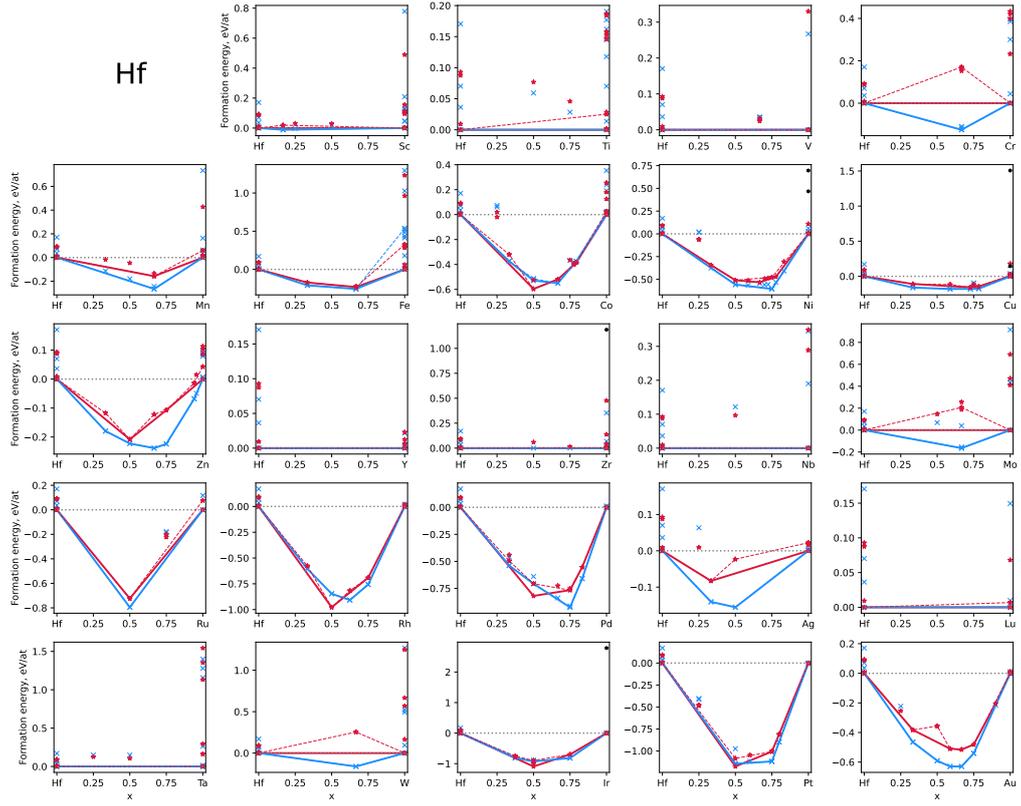

**Hf**

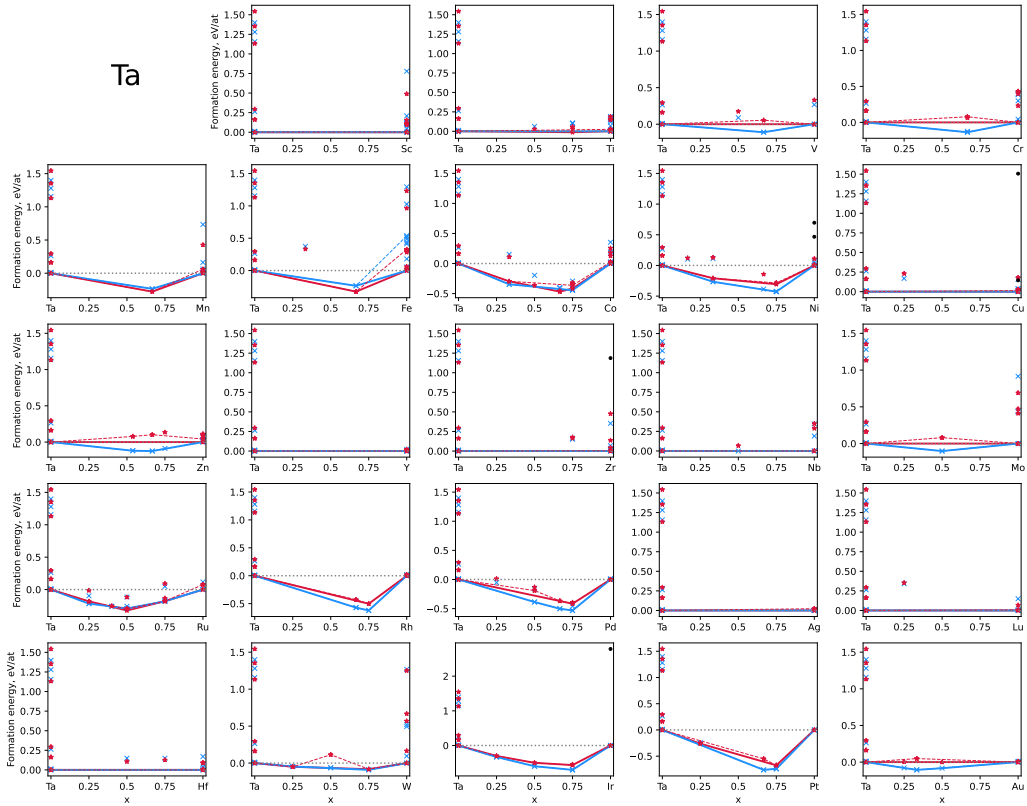

**Ta**

W

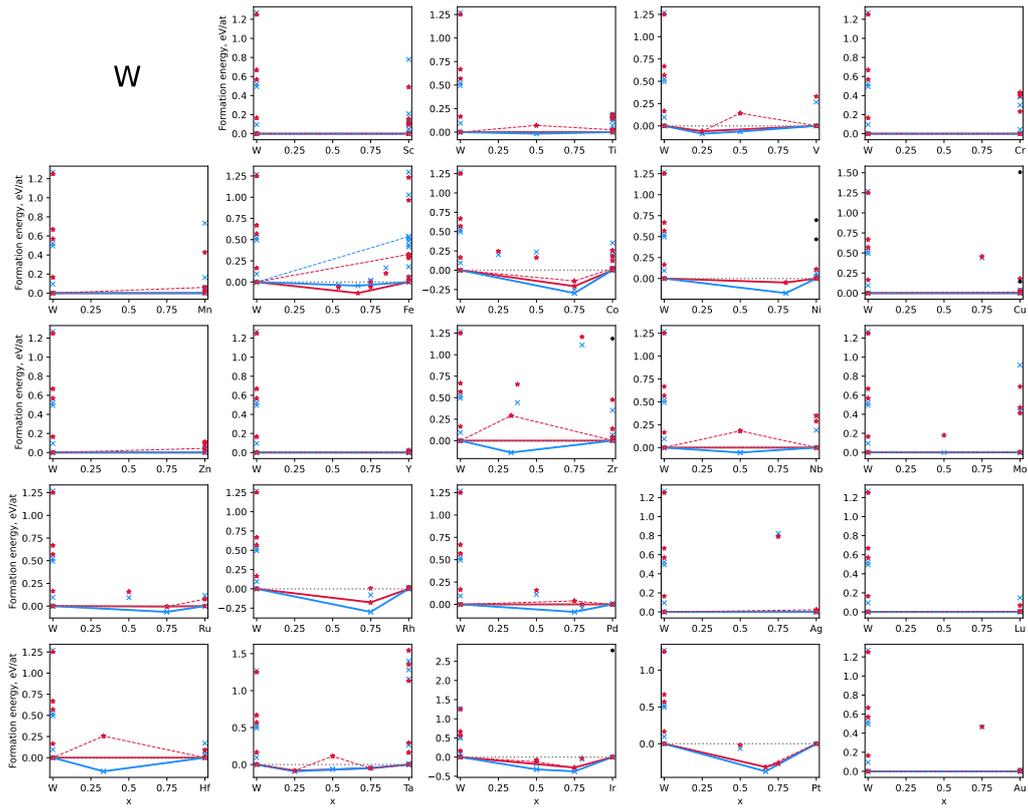

Ir

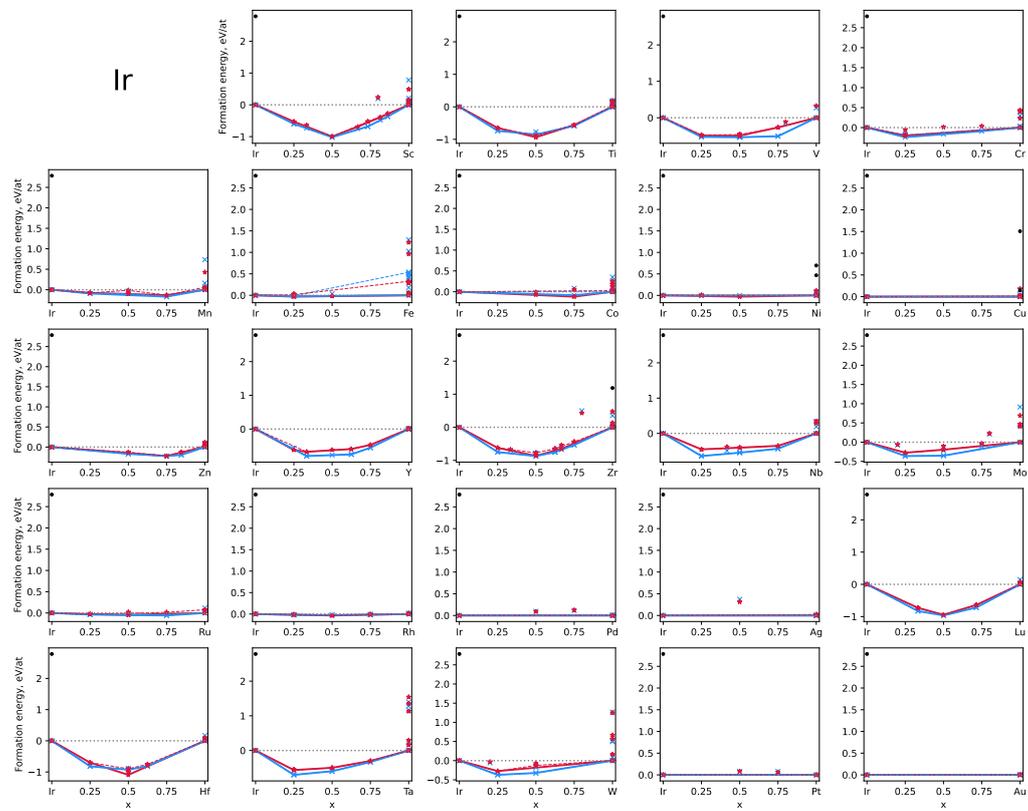

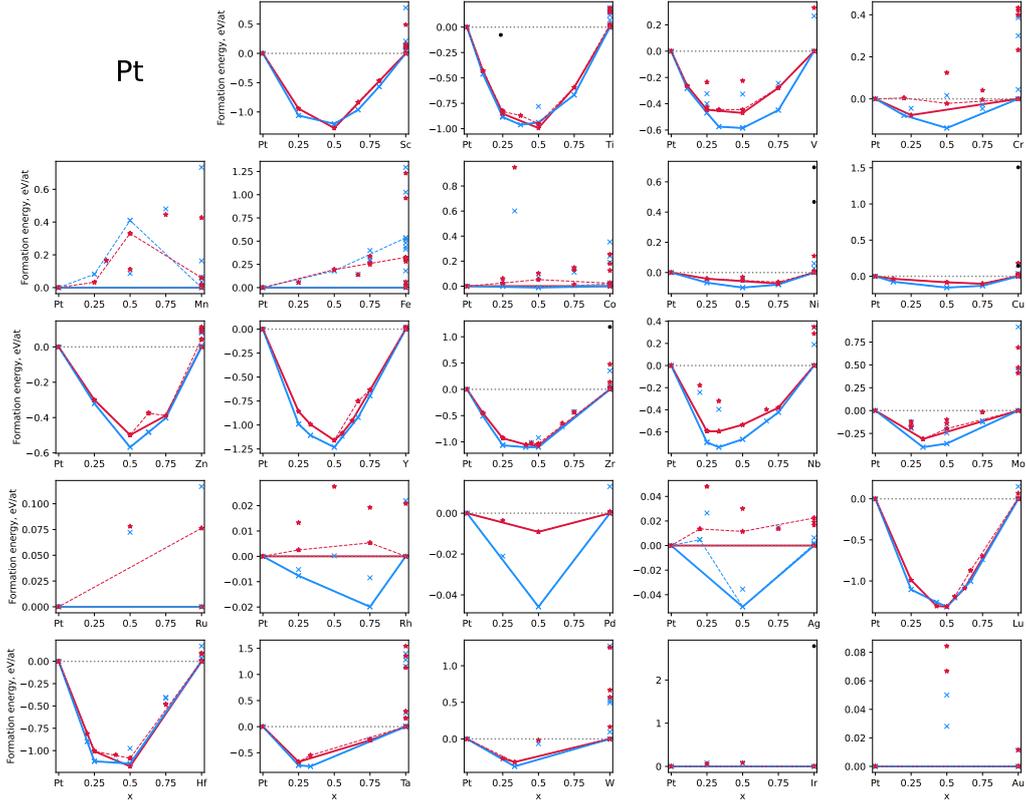

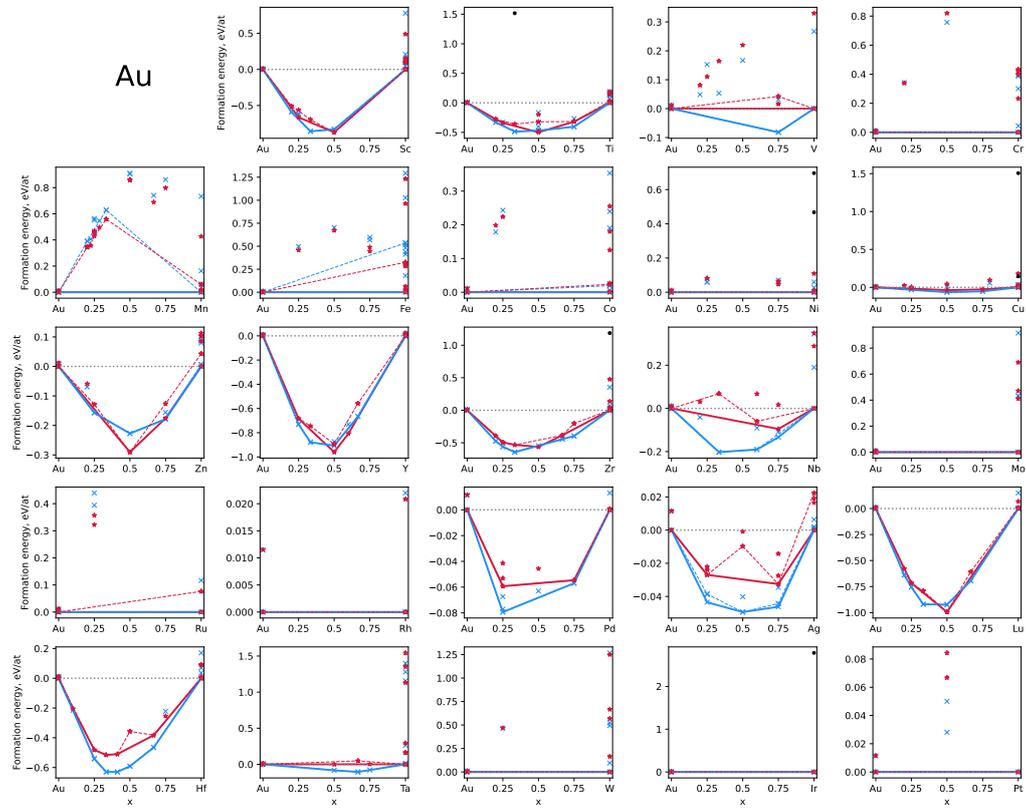


\end{document}